\newcommand\Msun{\hbox{M$_\odot$}}
\newcommand\Zsun{\hbox{Z$_\odot$}}
\newcommand\kms{\hbox{$\,$km$\,$s$^{-1}$}}

\newcommand\one{\,{\sc i}}
\newcommand\two{\,{\sc ii}}
\newcommand\three{\,{\sc iii}}
\newcommand\four{\,{\sc iv}}
\newcommand\five{\,{\sc v}}
\newcommand\tmult{\multicolumn{2}{c}}
\newcommand\hst{\textit{HST}}
\documentclass[12pt,preprint]{emulateapj}
\usepackage{times}
\usepackage{graphicx}
\usepackage{color}
\shorttitle{A spectroscopic census of M82 clusters}
\shortauthors{I. S. Konstantopoulos et al.}
\begin{document}
\title{A Spectroscopic Census of the M82 Stellar Cluster population$^{1,2}$}
\author{I. S. Konstantopoulos\altaffilmark{3}}
\author{N. Bastian\altaffilmark{4,3}}
\author{L. J. Smith\altaffilmark{5,3}}
\author{M. S. Westmoquette\altaffilmark{3}}
\author{G. Trancho\altaffilmark{6,7}}
\author{J. S. Gallagher III\altaffilmark{8}}

\altaffiltext{1}{Based on observations obtained at the Gemini Observatory, which is operated by the Association of Universities for Research in Astronomy, Inc., under a cooperative agreement with the NSF on behalf of the Gemini partnership: the National Science Foundation (United States), the Science and Technology Facilities Council (United Kingdom), the National Research Council (Canada), CONICYT (Chile), the Australian Research Council (Australia), CNPq (Brazil) and CONICET (Argentina).}
\altaffiltext{2}{Based on observations made with the NASA/ESA Hubble Space 
Telescope, obtained at the Space Telescope 
Science Institute, which is operated by the Association of Universities for 
Research in Astronomy, Inc., under NASA contract NAS5-26555. These observations are associated with program \#10853}
\altaffiltext{3}{Department of Physics and Astronomy, University College London, Gower Street, London, WC1E 6BT, UK; isk@star.ucl.ac.uk}
\altaffiltext{4}{Institute of Astronomy, University of Cambridge, Madingley Road, Cambridge}
\altaffiltext{5}{Space Telescope Science Institute and European Space Agency, 3700 San Martin Drive, Baltimore, MD 21218, USA}
\altaffiltext{6}{Universidad de La Laguna, Avenida Astr\'{o}fisico Francisco S\'{a}nchez s/n, 38206, La Laguna, Tenerife, Canary Islands, Spain}
\altaffiltext{7}{Gemini Observatory, 670 N. A'ahoku Place, Hilo, HI 96720, USA}
\altaffiltext{8}{Department of Astronomy, University of Wisconsin-Madison, 5534 Sterling, 475 North Charter Street, Madison, WI 53706, USA}
\clearpage
\begin{abstract}
We present a spectroscopic study of the stellar cluster population of M82, the archetype starburst galaxy, based primarily on new Gemini-North multi-object spectroscopy of 49 star clusters. These observations constitute the largest to date spectroscopic dataset of extragalactic young clusters, giving virtually continuous coverage across the galaxy; we use these data to deduce information about the clusters as well as the M82 post-starburst disk and nuclear starburst environments. Spectroscopic age-dating places clusters in the nucleus and disk between (7, 15) and (30, 270)~Myr, with distribution peaks at $\sim10$ and $\sim140$~Myr respectively. We find cluster radial velocities in the range \mbox{$v_\textup{\scriptsize R} \in (-160, 220)\,\kms$} ({\it wrt} the galaxy centre) and line of sight Na\,I~D interstellar absorption line velocities \mbox{$v_\textup{\scriptsize R}^\textup{\scriptsize Na\,I~D} \in (-75, 200)\,\kms$}, in many cases entirely decoupled from the clusters. As the disk cluster radial velocities lie on the flat part of the galaxy rotation curve, we conclude that they comprise a regularly orbiting system. Our observations suggest that the largest part of the population was created as a result of the close encounter with M81 $\sim220$~Myr ago. Clusters in the nucleus are found in solid body rotation on the bar. The possible detection of WR features in their spectra indicates that cluster formation continues in the central starburst zone. We also report the potential discovery of two old populous clusters in the halo of M82, aged $\gtrsim8$~Gyr. Using these measurements and simple dynamical considerations, we derive a toy model for the invisible physical structure of the galaxy, and confirm the existence of two dominant spiral arms.
\end{abstract}

\section{Introduction}\label{sec:intro}
The evolution of galaxies and their star formation rates are closely linked to interactions with other systems. The M81 group of galaxies provides a good local laboratory for studying such interactions and the ensuing violent episodes of star formation. The irregular galaxy M82 is believed to have suffered a close encounter with M81 $\sim220$ Myr ago (Brouillet et al. 1991; Yun, Ho \& Lo 1994; Yun 1999). This interaction evidently triggered the powerful starburst event that is currently seen in the central region. The first systematic study of the central starburst was undertaken by \citet{om78} who identified a number of optically high surface brightness starburst clumps (lettered A--E). These have subsequently each been found to contain hundreds of super star clusters \citep[SSCs;][]{oconnell95, melo05}. Unfortunately, however, the nearly edge-on inclination of the galaxy and its high dust content mean that, in projection, M82 presents a highly irregular, confused morphology with optical obscuration levels reaching 10--15~mags \citep{satyapal95, alonso03}. 

Studies of star clusters outside of the central regions have been relatively scarce because of the high levels of obscuration. \citet{GnS} and \citet{SnG} studied the isolated star cluster M82-F using a combination of \textit{HST} imaging and ground-based spectroscopy. The derived age and mass of $60\pm 20$~Myr and $1.2\pm 0.1\times 10^{6}$~\Msun\ \citep[revisited in][]{bastian07m82f} for M82-F show that 
significant star formation events have occurred in the M82 disk before the current nuclear starburst. \citet{RdG01} studied the cluster population of region B \citep[a $\sim1$~kpc region in the eastern part of the disk; see also][]{RdG03a}. Using optical $BVI$ photometry measured from \textit{HST}/WFPC2 imaging, they found that the region B clusters have ages of 0.5--1.5 Gyr, leading to region B being dubbed a `fossil starburst' region. However, not only are optical studies significantly affected by high levels of extinction, there also exists a degeneracy between the integrated optical light properties of young ($\lesssim$300~Myr) and old ($\gtrsim$1~Gyr) clusters when only using $BVI$ photometry, since the baseline provided does not include the age-sensitive Balmer jump. This feature lies at $\sim3700~$~\AA\ and is therefore `shared' between the $U$ and $B$ bands. It is thus difficult to break this degeneracy without imaging in at least one additional filter \citep{anders04}. This is due to the entanglement of the age and extinction measurements that have to be performed simultaneously when using photometry. 
In the absence of $U$-band coverage, the simple stellar population (SSP) model tracks to which we compare overlap, giving rise to a number of solutions -- often vastly different with respect to age. The inclusion of $U$-band goes a long way towards breaking this degeneracy, producing a single solution, i.~e.~ a single pair of age and extinction values. In our recent studies of region~B \citep[][hereafter Paper I]{smith07regb,isk08a} we used \textit{HST}/ACS \textit{UBVI} photometry, which covers the Balmer jump region, and Gemini-GMOS spectroscopy, where no degeneracy is present. Based on these data we showed that the region~B clusters are much younger than previously reported, with ages peaking at $\sim 150$~Myr, and that region B is optically bright because it represents a window into the body of the galaxy, where the foreground extinction is lower than in the surrounding areas.  A recent spectroscopic study by \citet{mayya06} also found the stellar population of the M82 disk to be consistent with having formed in the last Gyr, during a single, short ($\sim0.3$~Gyr) burst of star formation. 

Cluster spectra are affected to a lesser extent by extinction, and provide more reliable age determinations, as well as access to a wealth of other information (e.g.\ velocity and metallicity)---albeit at the cost of observing time \citep[e.g.][]{gelys07a}. In this paper we present optical spectroscopy for 49 clusters in M82, comprising the largest spectroscopic sample to date of young extragalactic star clusters. The observations were planned in such a way as to provide not only a statistically viable sample of the post-starburst cluster population, but almost continuous coverage across the disk and nucleus. This enables us to study the star clusters individually and as a population, and also infer their parameters as a function of the M82 starburst environment.

In this study we have sought to establish an age-dating technique based on the fit between observation and a range of models, that is as free as possible from the effects of extinction (which, as alluded to above, is a significant obstacle in M82). To diagnose the cluster properties, the main tool we employ is the optimised fitting of evolutionary synthesis models (in this case SSPs) to both the optical cluster spectral energy distributions (SEDs), and to individual Balmer absorption lines. This automatically breaks the aforementioned degeneracy between young/old ages, since the shape of the Balmer absorption lines is very sensitive to age. Furthermore, effects due to emission from foreground gas can be mitigated by careful masking of the line profile.

This paper is organised as follows: we present the imaging and spectroscopic datasets in \S~2; the measurements of ages and radial velocities from spectroscopy, and colours and reddening from photometry are presented in \S~3; we discuss the implications of these results in \S~4; and propose a model for the obscured physical structure of M82 in \S~5. We summarise all our results in the final section, \S~6. We adopt a distance to M82 of 3.6 Mpc (Freedman et al. 1994).

\section{Data acquisition and reduction}
\subsection{The full M82 spectroscopic dataset}
\begin{figure*}
\begin{center}
\plotone{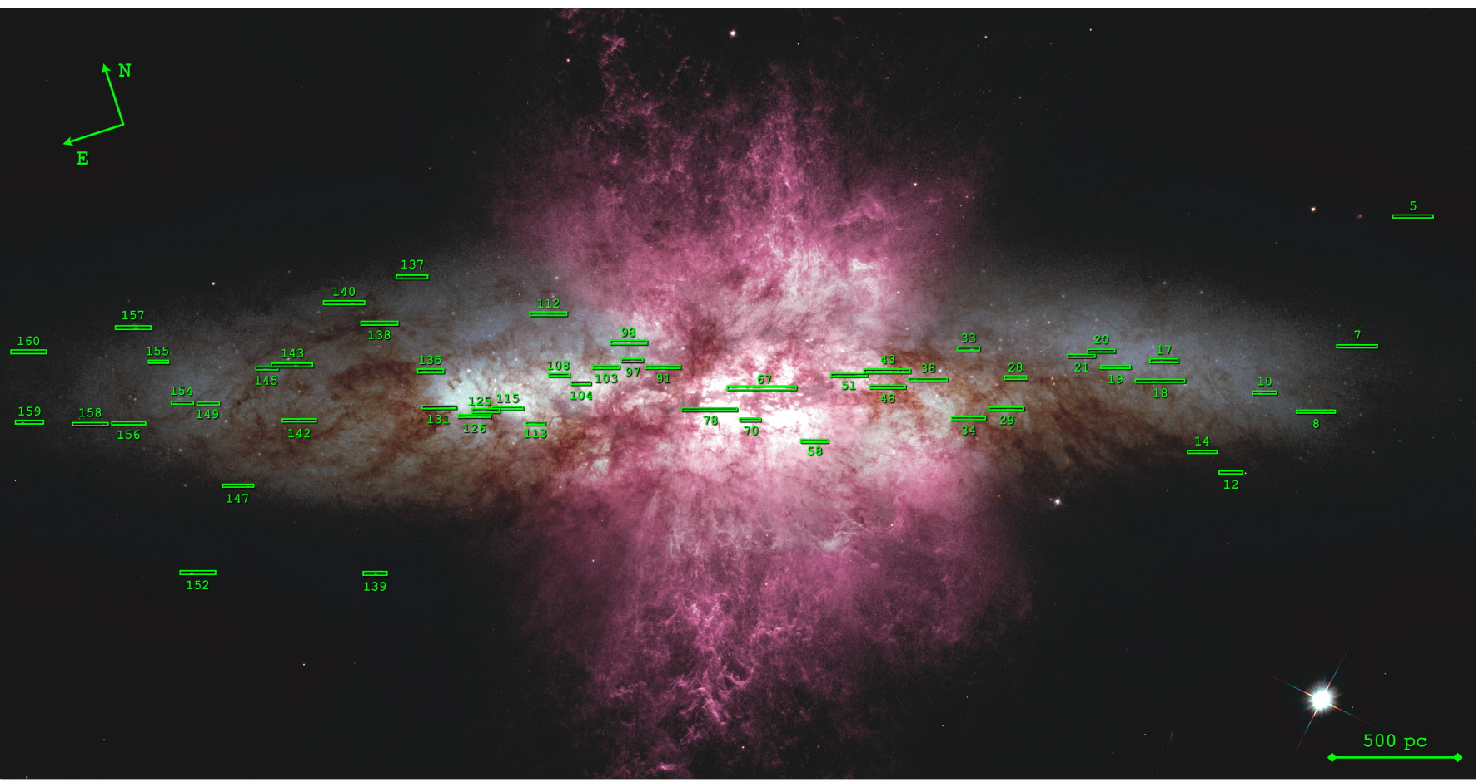}
\plotone{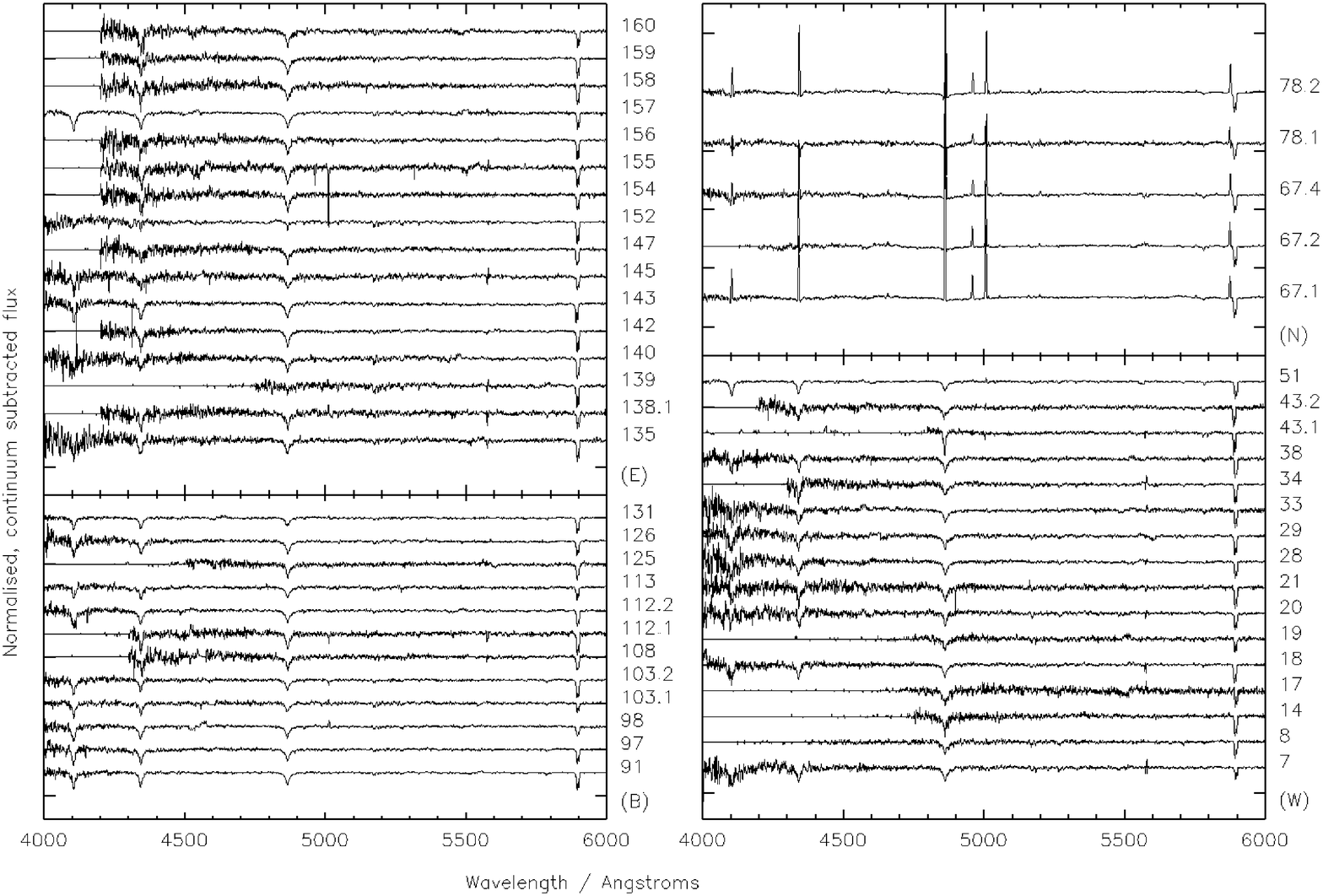}
\caption{The {\bf top panel} shows the existing dataset of star cluster spectroscopy (programmes Q38 and Q21, 16 hours of GMOS-MOS on-target integration time in all), with slit positions plotted over {\it HST}-ACS imaging of M82 \citep[Hubble Heritage ACS mosaic of 8 WFC images, ][]{mutchler07}. The dataset samples individual star clusters across the entire galaxy disk and cluster associations in the nucleus. In the {\bf bottom panel} we present all spectra used in this study, grouped according to the region the clusters occupy: `W', `B', `N' and `E' stand for western disk, region~B, nuclear region and eastern disk respectively. Note that we have excluded spectral regions with ${\rm S/N}<20$ from the plot. 
}
\label{plot:m82}
\end{center}
\end{figure*}

The results presented in this paper are based on spectroscopy obtained with the Gemini-North Multi-Object Spectrograph (GMOS-N) on the Gemini-North telescope, as part of observing program GN-2006A-Q38 and the follow-up program GN-2007A-Q21 (PI L.~J.~Smith; hereafter Q38 and Q21). Overall, two multi-object slit masks were designed and included a total of 54 slits, from which we extracted 62 spectra in the range $\lambda\lambda3700$--6500\,\AA. Figure~\ref{plot:m82} marks the positions of the slits on \hst/ACS (Advanced Camera for Surveys) imaging of M82 \citep{mutchler07} and presents all the cluster spectra used in the subsequent analysis.

The source selection was performed within the bounds of the original observing program (Q38). The candidate young massive clusters (YMCs) were selected based on GMOS pre-imaging after cross-identification with \hst~imaging, with the aim of choosing the brightest isolated clusters, spanning most of the optical extent of the disk. Starting from a masterlist of several hundred candidates, we built two multi-slit masks (one for each observing program), with a standard $0\farcs75$ slit width, while the slit length was varied to optimize the overall arrangement. In all cases, we attempted to apply a minimum slit length of $6\farcs5$, in order to provide adequate sky for background subtraction. In Q38, we fell below $6\farcs5$ in a few cases, to accommodate more clusters along the slit; this, however, complicated the aperture extraction procedure. For that reason, in Q21 we did not sacrifice the slit length in favour of cluster number, particularly since the mean magnitude was fainter compared to Q38. 

The GMOS observational setup was identical for Q38 and Q21. We used the B600 grating ($0.9$~\AA\,px$^{-1}$) and the three GMOS CCDs in $2\times2$ binning mode. 
The observations were executed in pairs of pointings with different central wavelengths (5080~\AA\ and 5120~\AA) in order to cover the inter-chip gaps.
A brief journal of observations is given in Table~\ref{tab:obs}. The total exposure time for each dataset was eight 1800~s exposures for a total on-source integration time of 4 hr. CuAr arc exposures and Quartz-Halogen flat fields were taken in between target exposures, and bias frames were taken as part of the Gemini base-line calibrations (GCAL). 

The data reduction was performed using standard IRAF\footnote{
IRAF is distributed by the National Optical Astronomical Observatories, which are operated by the Association of Universities for Research in Astronomy, Inc. under contract with the National Science Foundation} 
and purpose-designed Gemini-IRAF routines. In brief, this consisted of bias-subtraction, flat-fielding, and the combination of a three-CCD image into a single frame, while correcting for the variation in detector quantum efficiency. We used the obtained CuAr frames for the wavelength calibration, which resulted in satisfactorily low residuals (typically less than 0.2~\AA). As Gemini does not have a functioning atmospheric diffraction corrector, we corrected for differential atmospheric refraction using a purpose-built IDL (Interactive Data Language) routine written by B.~W.~Miller \citep[private communication; this routine is based on][where the effects of atmospheric differential refraction are discussed]{filippenko82}. Finally, all eight available exposures of each target were extracted and then
flux calibrated using exposures of standard stars Wolf~1346 (observed 2006 May 25) and {HZ~44} (observed 2008 Feb 18) for Q38 and Q21 respectively.

\begin{figure}
\begin{center}
\plotone{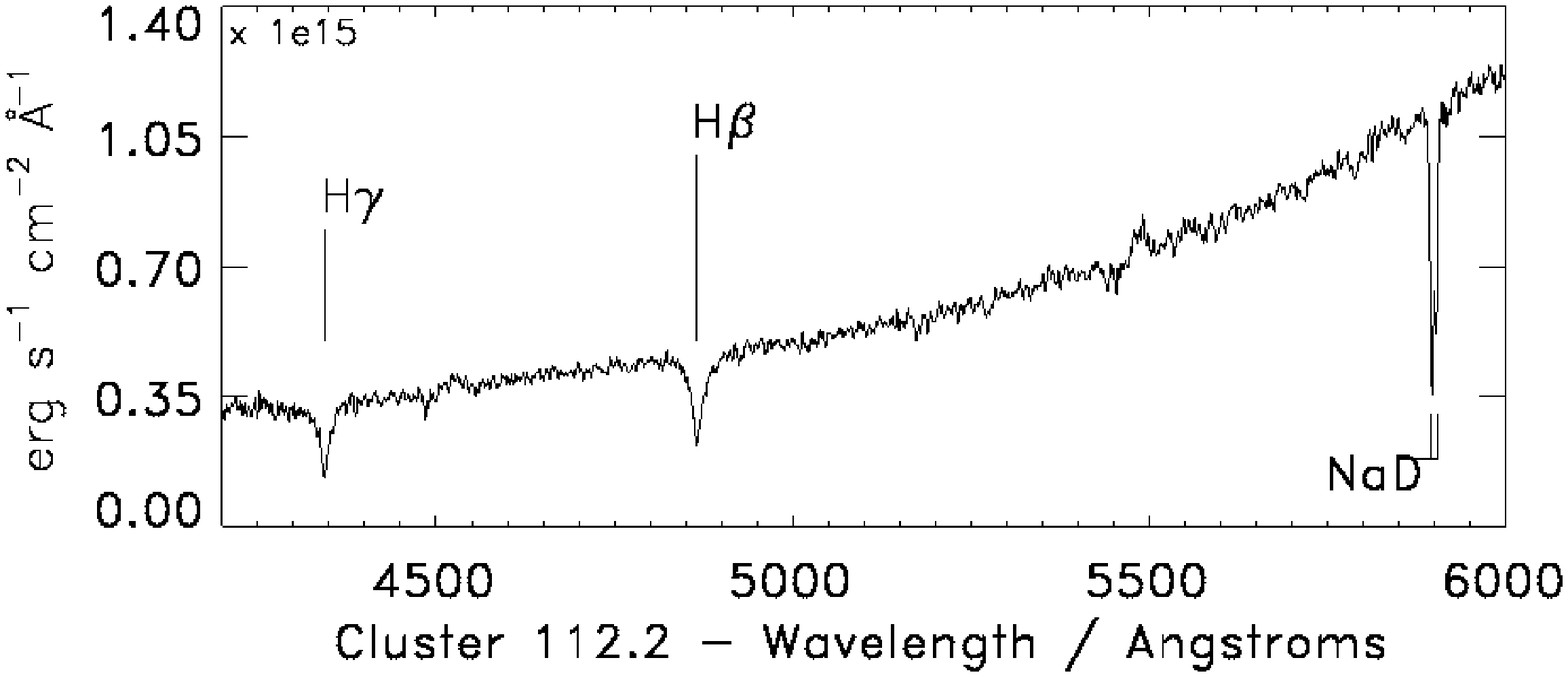}
\caption{A representative flux-calibrated, high S/N spectrum, with the main absorption features denoted.}
\label{plot:example}
\end{center}
\end{figure}

In summary, the two masks contained 54 slits (see Table~\ref{tab:obs}) and the total number of sources observed was 62, of which 28 have a S/N above 20 (conservatively measured in two continuum windows covering $\lambda\lambda$4500--4800 \AA\ and $\lambda\lambda$5000--5500 \AA). We present spectra for 49 clusters in this paper, and omit the remaining 13 on grounds of low data quality. The spectral resolution of the Q38 and Q21 cluster sample was measured to be 3.5~\AA\ and 3.7~\AA\ respectively from the CuAr spectra, but we assume a common resolution of 3.7~\AA\ for the sake of simplicity. The spectral range of our data is $\approx \lambda\lambda$3700--6500~\AA, which includes the H$\beta$ and H$\gamma$ lines, our main age indicators, and the Na\,I\, D doublet, which we use as a tracer of interstellar gas radial velocities. The exact wavelength range varies with each slit, depending on its positioning with respect to the M82 major axis, which defines the dispersion axis. Our effective wavelength range for all slits is $\lambda\lambda$4000--6000~\AA; this is set by the poor sensitivity of GMOS below 4000~\AA\ and an uncertain flux calibration above 6000~\AA. { Fig.~\ref{plot:example} demonstrates these features on the representative high S/N spectrum of cluster~112.2}. Positional and other basic information for all clusters can be found in Table~\ref{tab:phot} (the coordinates of our sources have been converted to the \hst~WCS, as the mosaic imaging is publicly available).

The cataloging convention we have adopted for cluster naming follows the increasing slit number in the masterlist compiled during the initial source selection. The list is not continuous (as more sources were initially selected than could be accommodated), but the increasing number does indicate the position of a cluster along the M82 disk -- numbers count from west to east, along the major axis. In addition, several slits include multiple sources; in these cases, we refer to clusters by the slit number with a decimal part, to indicate the increasing number of the source (again, counting from west to east).
Our sample spans most of the extent of the optically visible disk, which is approximately 6~by~2~kpc, as seen in \hst~imaging. 

\begin{table}
\caption{Summary of spectroscopic observations taken with Gemini-North}
\label{tab:obs}
\begin{center}
\begin{tabular}{lcccc}
\tableline
\tableline
Dataset	&	Dates of observation	&	Seeing\tablenotemark{a}	&No. slits	&	Exposure time \\
\tableline
GN-2006A-Q38 & 2006 April 5		&	$0\farcs8$	&	39	& 	$8\times1800$~s\\
GN-2007A-Q21 & 2007 June 4--8	&	$0\farcs45$\tablenotemark{b}	&	15	& 	$8\times1800$~s\\
\tableline
\end{tabular}
\end{center}
\tablenotetext{a}{This is the Gemini `image-quality' (IQ) at 5000\AA. This parameter accounts for wind and telescope performance effects as well as atmospheric seeing.}
\tablenotetext{b}{For all nights except 2007 June 08, when the  IQ was 0\farcs8.}
\end{table}

\begin{table*}
\caption{Positional information (equinox J2000) and aperture photometry for all 49 clusters in the spectroscopic sample. 
}\scriptsize
\label{tab:phot}
\begin{center}
\begin{tabular}{lccrrcrrlrlrlccl}
\tableline
\tableline
ID (alt ID)	&	\multicolumn{3}{c}{RA}		&	\multicolumn{3}{c}{~~~~~DEC}	& ~$m_B$	&   $\pm$ &	$m_V$	&   ~~~$\pm$ &	$m_I$	&   $\pm$ &   $A_V$&  $M_V$&   ~~~$\pm$\\
	&	(h	&	m	&	s)		&	\hspace{0.5cm}~($^\circ$&$\prime$&$\prime\prime$)		&	\multicolumn{2}{c}{(mag)}			&	\multicolumn{2}{c}{(mag)}			&		\multicolumn{2}{c}{(mag)}		&  (mag)  &\tmult{(mag)}\\
\tableline
    7	&     9 &    55 &  27.53 &    69 &    40 &    3.7 &   21.02 &	 0.09 &   20.55 &    0.07 &   19.67 &	 0.05 &    0.85 &   -8.09 &    0.95 \\
    8	&     9 &    55 &  29.99 &    69 &    39 &   53.6 &   21.22 &	 0.10 &   20.70 &    0.08 &   19.79 &	 0.06 &    0.88 &   -7.96 &    0.91 \\
   14	&     9 &    55 &  35.46 &    69 &    39 &   56.5 &   21.93 &	 0.13 &   20.89 &    0.08 &   19.48 &	 0.05 &    2.03 &   -8.92 &    0.44 \\
   17	&     9 &    55 &  35.28 &    69 &    40 &   18.5 &   20.70 &	 0.08 &   20.08 &    0.06 &   19.08 &	 0.05 &    1.10 &   -8.80 &    0.80 \\
   18	&     9 &    55 &  35.99 &    69 &    40 &   14.9 &   20.37 &	 0.07 &   19.76 &    0.06 &   18.74 &	 0.04 &    1.15 &   -9.17 &    0.80 \\
   19	&     9 &    55 &  37.11 &    69 &    40 &   20.9 &   20.76 &	 0.08 &   20.24 &    0.06 &   19.39 &	 0.06 &    0.98 &   -8.51 &    0.87 \\
   20	&     9 &    55 &  37.50 &    69 &    40 &   26.0 &   21.25 &	 0.10 &   20.77 &    0.08 &   19.94 &	 0.06 &    0.75 &   -7.76 &    1.03 \\
   21	&     9 &    55 &  38.56 &    69 &    40 &   27.0 &   21.01 &	 0.09 &   20.31 &    0.07 &   19.41 &	 0.06 &    1.02 &   -8.49 &    0.83 \\
   28	&     9 &    55 &  41.03 &    69 &    40 &   27.8 &   21.27 &	 0.10 &   20.46 &    0.07 &   19.16 &	 0.04 &    1.70 &   -9.02 &    0.53 \\
   29	&     9 &    55 &  41.85 &    69 &    40 &   22.1 &   21.07 &	 0.10 &   20.07 &    0.06 &   18.55 &	 0.04 &    2.17 &   -9.88 &    0.46 \\
   33	&     9 &    55 &  42.40 &    69 &    40 &   37.8 &   21.54 &	 0.11 &   20.26 &    0.06 &   18.46 &	 0.03 &    2.80 &  -10.32 &    0.37 \\
   34	&     9 &    55 &  43.69 &    69 &    40 &   23.4 &   21.11 &	 0.09 &   19.82 &    0.06 &   18.14 &	 0.04 &    2.62 &  -10.58 &    0.40 \\
   38	&     9 &    55 &  45.12 &    69 &    40 &   36.6 &   20.97 &	 0.09 &   20.20 &    0.06 &   19.09 &	 0.05 &    1.42 &   -9.00 &    0.63 \\
   43.1 &     9 &    55 &  45.61 &    69 &    40 &   40.1 &   20.63 &	 0.08 &   19.14 &    0.05 &   17.13 &	 0.02 &    3.22 &  -11.87 &    0.37 \\
   43.2 &     9 &    55 &  46.62 &    69 &    40 &   43.9 &   19.90 &	 0.06 &   19.27 &    0.05 &   18.35 &	 0.05 &    1.02 &   -9.54 &    0.93 \\
   51 (F\tablenotemark{a}) & 9 &    55 &  47.08 &    69 &    40 &   42.3 &   17.61 &	 0.02 &   16.50 &    0.01 &   14.94 &	 0.01 &    2.42 &  -13.70 &    0.57 \\
   58	&     9 &    55 &  50.27 &    69 &    40 &   35.0 &   20.25 &	 0.08 &   19.72 &    0.07 &   18.75 &	 0.07 &    1.12 &   -9.19 &    0.82 \\
   67.1 &     9 &    55 &  50.44 &    69 &    40 &   47.2 &   19.92 &	 0.05 &   18.72 &    0.04 &   17.04 &	 0.05 &    2.47 &  -11.54 &    0.47 \\
   67.2 &     9 &    55 &  50.77 &    69 &    40 &   47.9 &   21.18 &	 0.13 &   20.69 &    0.26 &   18.78 &	 0.15 &    2.17 &   -9.27 &    0.44 \\
   67.4 &     9 &    55 &  51.93 &    69 &    40 &   50.2 &   21.60 &	 0.12 &   20.80 &    0.11 &   19.12 &	 0.22 &    2.12 &   -9.11 &    0.43 \\
   78.1 &     9 &    55 &  52.64 &    69 &    40 &   46.9 &   21.16 &	 0.14 &   20.76 &    0.23 &   19.50 &	 0.40 &    1.20 &   -8.22 &    0.69 \\
   78.2 &     9 &    55 &  53.13 &    69 &    40 &   49.1 &   20.43 &	 0.10 &   19.36 &    0.08 &   16.89 &	 0.07 &    3.45 &  -11.87 &    0.35 \\
   91 (H\tablenotemark{b}) &     9 &    55 &  54.61 &    69 &    41 &    1.6 &   19.20 &	 0.04 &   18.03 &    0.02 &   16.44 &	 0.01 &    2.40 &  -12.15 &    0.51 \\
   97	&     9 &    55 &  55.65 &    69 &    41 &    5.9 &   19.08 &	 0.04 &   18.56 &    0.03 &   17.63 &	 0.02 &    0.90 &  -10.12 &    1.12 \\
   98	&     9 &    55 &  55.50 &    69 &    41 &    9.9 &   19.69 &	 0.05 &   19.31 &    0.04 &   18.58 &	 0.04 &    0.50 &   -8.97 &    1.79 \\
  103.1 &     9 &    55 &  56.47 &    69 &    41 &    6.4 &   20.62 &	 0.09 &   20.35 &    0.10 &   20.10 &	 0.20 &    0.00 &   -7.43 &    1.00 \\
  103.2 &     9 &    55 &  57.04 &    69 &    41 &    7.3 &   19.85 &	 0.05 &   19.10 &    0.04 &   17.88 &	 0.03 &    1.58 &  -10.26 &    0.65 \\
  108	&     9 &    55 &  58.43 &    69 &    41 &    8.7 &   20.66 &	 0.10 &   19.81 &    0.07 &   18.27 &	 0.04 &    2.03 &  -10.00 &    0.50 \\
  112.2 &     9 &    55 &  58.26 &    69 &    41 &   23.2 &   20.71 &	 0.08 &   20.21 &    0.06 &   19.41 &	 0.05 &    0.70 &   -8.27 &    1.18 \\
  113	&     9 &    56 &   0.63 &    69 &    41 &    2.5 &   20.19 &	 0.09 &   19.66 &    0.08 &   18.19 &	 0.04 &    1.67\tablenotemark{c} &   -9.79 &    0.59 \\
  125	&     9 &    56 &   2.28 &    69 &    41 &    9.0 &   19.46 &	 0.05 &   18.79 &    0.04 &   17.77 &	 0.03 &    1.17 &  -10.16 &    0.87 \\
  126	&     9 &    56 &   2.53 &    69 &    41 &    8.1 &   19.22 &	 0.04 &   18.65 &    0.03 &   17.79 &	 0.03 &    0.85 &   -9.98 &    1.17 \\
  131	&     9 &    56 &   3.40 &    69 &    41 &   12.2 &   18.51 &	 0.03 &   17.93 &    0.03 &   16.94 &	 0.02 &    1.08 &  -10.93 &    1.02 \\
  135	&     9 &    56 &   3.87 &    69 &    41 &   22.4 &   21.02 &	 0.09 &   20.71 &    0.09 &   19.28 &	 0.05 &    1.45\tablenotemark{d} &   -8.52 &    0.59 \\
  138.1 &     9 &    56 &   4.49 &    69 &    41 &   35.3 &   22.30 &	 0.16 &   21.73 &    0.12 &   20.71 &	 0.09 &    1.25 &   -7.30 &    0.59 \\
  139	&     9 &    56 &   9.24 &    69 &    40 &   45.7 &   22.15 &	 0.14 &   20.44 &    0.07 &   18.30 &	 0.03 &    2.22 &   -9.57 &    0.43 \\
  140	&     9 &    56 &   5.51 &    69 &    41 &   42.2 &   21.09 &	 0.09 &   20.33 &    0.06 &   19.01 &	 0.04 &    1.90 &   -9.35 &    0.49 \\
  142	&     9 &    56 &   9.60 &    69 &    41 &   23.5 &   20.39 &	 0.06 &   19.60 &    0.05 &   18.40 &	 0.03 &    1.55 &   -9.74 &    0.63 \\
  143	&     9 &    56 &   8.72 &    69 &    41 &   35.0 &   20.83 &	 0.08 &   20.24 &    0.06 &   19.28 &	 0.04 &    1.05 &   -8.59 &    0.82 \\
  145	&     9 &    56 &   9.96 &    69 &    41 &   36.9 &   21.78 &	 0.12 &   21.29 &    0.10 &   20.44 &	 0.08 &    0.77 &   -7.26 &    0.94 \\
  147	&     9 &    56 &  12.66 &    69 &    41 &   14.7 &   21.41 &	 0.10 &   20.70 &    0.08 &   19.44 &	 0.05 &    1.65 &   -8.73 &    0.53 \\
  152	&     9 &    56 &  16.33 &    69 &    41 &    2.2 &   20.08 &	 0.06 &   18.96 &    0.03 &   17.48 &	 0.02 &    0.62 &   -9.45 &    1.51 \\
  154	&     9 &    56 &  14.06 &    69 &    41 &   37.9 &   20.58 &	 0.07 &   20.21 &    0.06 &   19.54 &	 0.06 &    0.40 &   -7.97 &    1.99 \\
  155	&     9 &    56 &  14.04 &    69 &    41 &   48.1 &   21.63 &	 0.11 &   21.22 &    0.10 &   20.55 &	 0.08 &    0.45 &   -7.01 &    1.56 \\
  156	&     9 &    56 &  15.71 &    69 &    41 &   36.6 &   21.35 &	 0.10 &   19.88 &    0.05 &   18.87 &	 0.04 &    4.56\tablenotemark{e} &   -9.80 &    0.52 \\
  157	&     9 &    56 &  14.26 &    69 &    41 &   56.7 &   20.52 &	 0.07 &   18.87 &    0.03 &   18.11 &	 0.03 &    5.13\tablenotemark{f} &  -10.59 &    0.63 \\
  158	&     9 &    56 &  17.18 &    69 &    41 &   40.0 &   19.28 &	 0.04 &   20.99 &    0.09 &   20.15 &	 0.07 &    0.00 &   -6.79 &    1.00 \\
  159	&     9 &    56 &  20.26 &    69 &    41 &   47.6 &   21.54 &	 0.11 &   20.82 &    0.08 &   19.90 &	 0.06 &    1.05 &   -8.01 &    0.76 \\
  160	&     9 &    56 &  19.26 &    69 &    42 &    2.5 &   21.60 &	 0.11 &   21.17 &    0.09 &   20.45 &	 0.08 &    0.55 &   -7.16 &    1.30 \\

\tableline
\end{tabular}
\end{center}
\tablenotetext{a,b}{Following the OM78 notation}
\tablenotetext{c,d,e,f}{These clusters are differentially extinguished across the three bands, and their $A_V$ measurements are therefore highly uncertain.}
\end{table*}

\subsection{HST Photometry}
We also make use of imaging data of M82 to derive cluster colours and deduce information about their surroundings. We use the publicly available Hubble Heritage ACS mosaic imaging  of M82 \citep[for a full description, we refer the reader to][]{mutchler07}, which covers the galaxy disk, nucleus and outskirts in six pointings {and in three filters, $F435W$, $F555W$, $F814W$. These are roughly equivalent to the standard Johnson BVI, so we use this notation throughout this paper. Note, however, that we do not perform any conversion between the two systems. All of our observed clusters lie within the field of view of the ACS imaging. The spatial scale of the images is $0.05$~arcsec~px$^{-1}$, and one pixel corresponds to $\sim0.9$~pc at the distance of M82.

We used the \hst~images to perform aperture photometry on all sources in our spectroscopic sample. For most disk clusters, we applied an aperture radius of 10 pixels to avoid contamination by nearby objects, with a 2 pixel wide sky annulus placed at 12 pixels. As most of our sources are of considerable size, we calculated aperture corrections for each filter as $-0.34$ mag for the $B$ and $V$~bands and $-0.45$~mag for the $I$~band, based on the flux difference between 10 and 30 pixel apertures for isolated clusters.  After close inspection of each source individually, it was deemed necessary to limit the aperture size in some cases, due to the presence of contaminating neighbouring sources. Thus, we applied 5~pixel apertures with the same annulus for eight disk clusters as well as all sources in the nuclear region (in the nucleus, the sky annulus was placed between 7 and 9~px), and calculated the corrections to 30~pixels as $-0.94$, $-0.90$ and $1.04$~mag for the $B$, $V$, and $I$ bands respectively. While these aperture corrections are significant and dependent on cluster size, the cluster colours should be largely unaffected.  The photometric results are presented in Table~\ref{tab:phot}; this includes cluster reddening, which is discussed in \S~\ref{sec:col}.

\section{Spectroscopic and photometric analysis}
\subsection{Cluster ages and radial velocities from spectroscopy}\label{sec:ages-rvs}
In this section we present our methodology for deriving cluster ages and illustrate our techinques through example fits. We employ the Cumulative $\chi^2$ Minimisation method (CCM) presented in Paper~I to obtain ages for the full sample of 49 star clusters in the disk and nuclear region of M82. This method is based on the goodness of fit between observed cluster spectra and evolutionary population synthesis (EPS) models. We chose to compare our spectra to the \citet[][hereafter GD05]{gd05} model library, a range of theoretical models specifically compiled for young star clusters. The major advantages these models present over more established empirical cluster SEDs \citep[e.~g.][]{bc03} are the high spectral and temporal resolution (0.3~\AA\ and a few Myr for the first 100~Myr). We used the full range of models (4~Myr to 17~Gyr) compiled using Padova isochrones, and a Salpeter IMF. We have assumed a solar metallicity following the findings of \citet{mcleod93} for the M82 ISM.  

Before the fit can take place, we rectify both observed spectrum and template (using two continuum windows adjacent to the absorption line) and smooth the models by a factor of three 
and then re-bin them to the wavelength scale of the observed spectra. In order to obtain a first estimate of the cluster age, we apply an approximate radial velocity (RV) correction on the observed spectra (matching the Balmer line profiles by eye; this calculation is later refined). We use the lower Balmer series line profiles (H$\beta$ and H$\gamma$) to perform the fits, and take care not to include any emission from ionised gas in the line of sight (which is plentiful in this gas rich, edge-on galaxy). This makes our age determination technique largely independent of the effects of foreground gas and dust. 

In brief, the CCM calculates the $\chi^2$ for each point in the fit, and picks the best fitting age as the model with the lowest `reduced' $\chi^{2}$,  expressed through $\chi^2_{\nu}={\Sigma\chi^2}/{d}$ (note that we will omit the subscript $\nu$ hereafter); $d$ is the number of degrees of freedom, taken as the number of points included in the fit. This approach follows the reasoning outlined in \citet{lampton76}. Our fitting technique is therefore sensitive to the overall line profile (core and wings) rather than just the line depth. This approach overcomes both the line strength degeneracy (discussed in \S~\ref{sec:intro}) that exists between models of age $\sim200\pm100$~Myr and $\sim600\pm100$~Myr, and the lack of information on the line core where the superimposed emission is strong or complex. We therefore obtain two measurements of the cluster age, based on the fits on the H$\gamma$ and  H$\beta$ features. As the two measurements do not always agree perfectly (usually due to the lower S/N in the H$\gamma$ spectral region), we examine each case individually. In a few cases, one of the two measurements is excluded from the fit; we apply this principle by enforcing a minimum acceptable S/N ratio of 10 (this applies to 12 clusters in the sample).

The uncertainty on a given fit is defined as the range of ages that lie within $1\sigma$ of the minimum $\chi^2$ -- quantified by the criterion $\chi^2_i\le \chi^2_{min} + N$, where $N$ is the number of free parameters in the fit \citep{lampton76}. In this case, the only varying parameter is the age of the model, therefore \mbox{$N=1$}. 
This process is slightly complicated by the fact that a large number of our adopted ages consist of the statistical mean of the results of two $\chi^2$ fits. In these cases, we combine the probability distributions derived for each fit (simply by adding them) and use a value of $N=2$ to define the $1\sigma$ confidence region. We stress that, while this error determination is not statistically flawless, it provides an  adequate description of the range of models that can plausibly describe the spectra. This range therefore matches what one might pick `by eye'. This statistical analysis was also successfully used in \citet{bastian08cores}.

Our fitting method is demonstrated in Fig.~\ref{plot:age-fits} for three clusters of varying luminosity and S/N. We present a $BVI$ colour-composite image (the spatial scale is a square of side 200~px or 175~pc) and the spectrum of each cluster, with the best-fit model overplotted. We also show the probability distributions of all available models as a function of age below each spectrum. We note that, although not demonstrated,  the H$\gamma$ and H$\beta$ fits do provide overlapping $\chi^2$ distributions (i.~e. equivalent fits). In the top panel we show a low S/N spectrum (cluster 19, ${\rm S/N}<20$), where the H$\beta$ region provides the only reliable fit. The middle panel shows the H$\gamma$ profile in a medium S/N ($\sim25$) spectrum. For this cluster, we are able to obtain two age measurements in good agreement. In the bottom panel we show the high S/N ($\sim40$) spectrum of cluster 112.2, where the H$\beta$ and H$\gamma$ measurements give the same age. {We note that these plots do not present a subset of high-quality fits, but are meant to provide an overview of all problems faced by our fitting method, be they due to low S/N, embedded/blended emission, or related to the template fitting process.}

\begin{figure*}
\begin{center}
\includegraphics[width=0.37\textwidth]{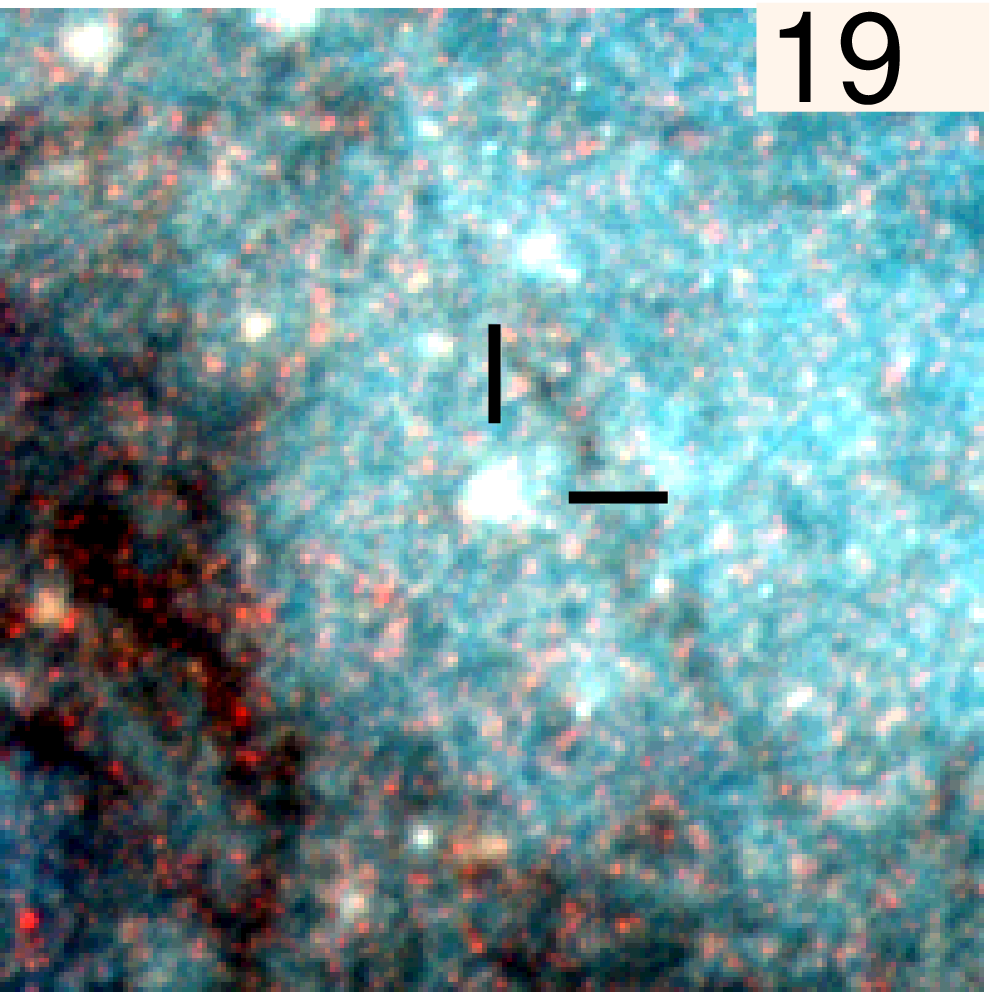}\hspace{5pt}\includegraphics[width=0.46\textwidth]{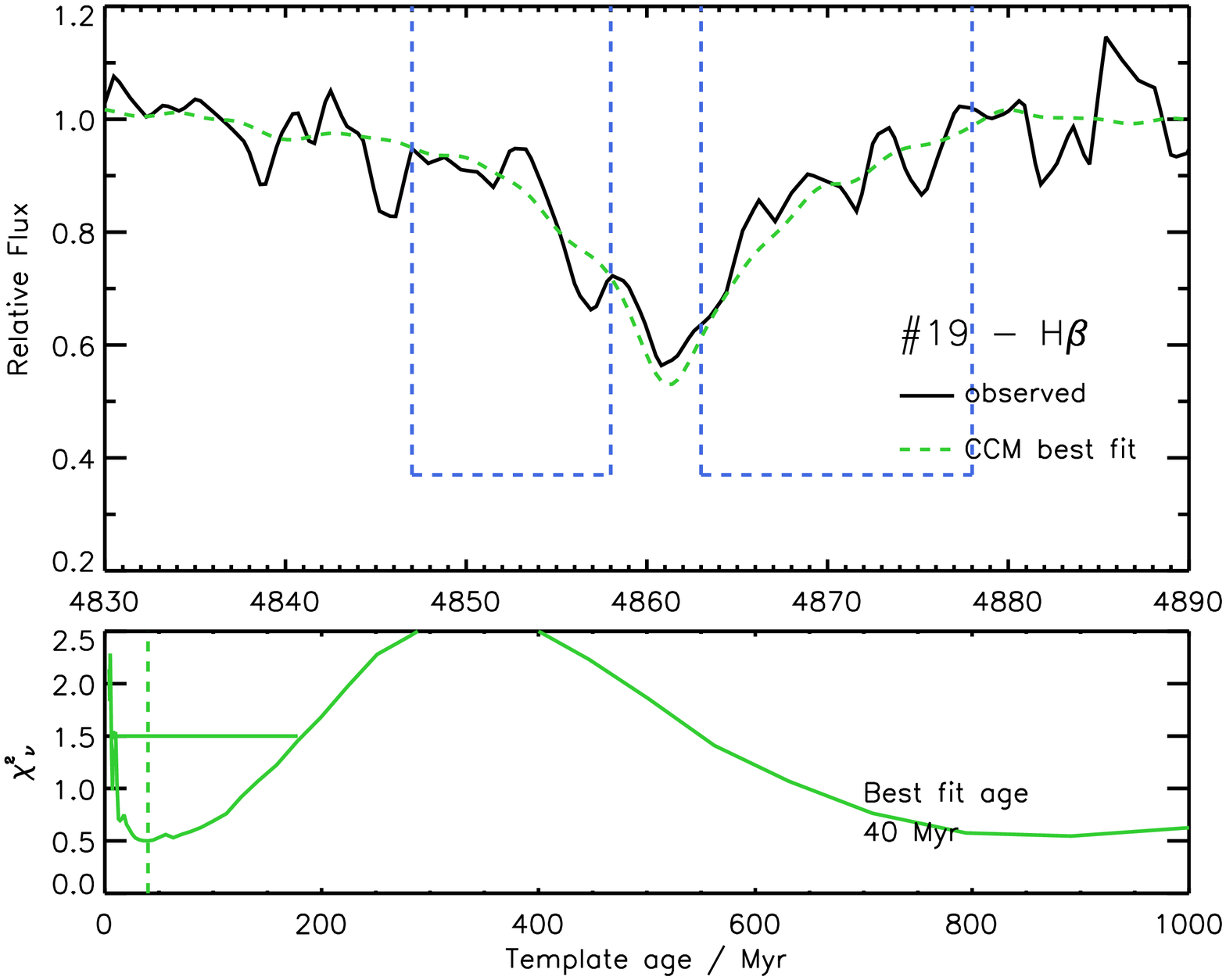}\vspace{5pt}
\includegraphics[width=0.37\textwidth]{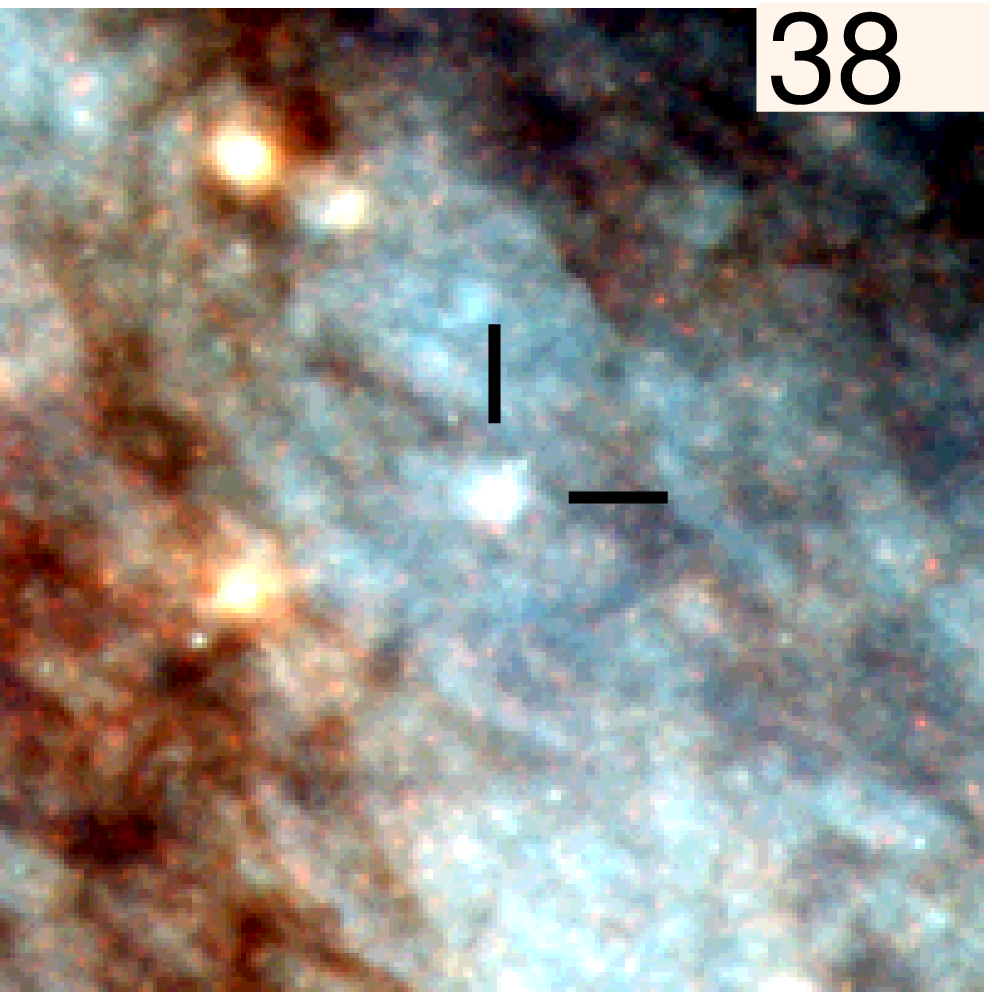}\hspace{5pt}\includegraphics[width=0.46\textwidth]{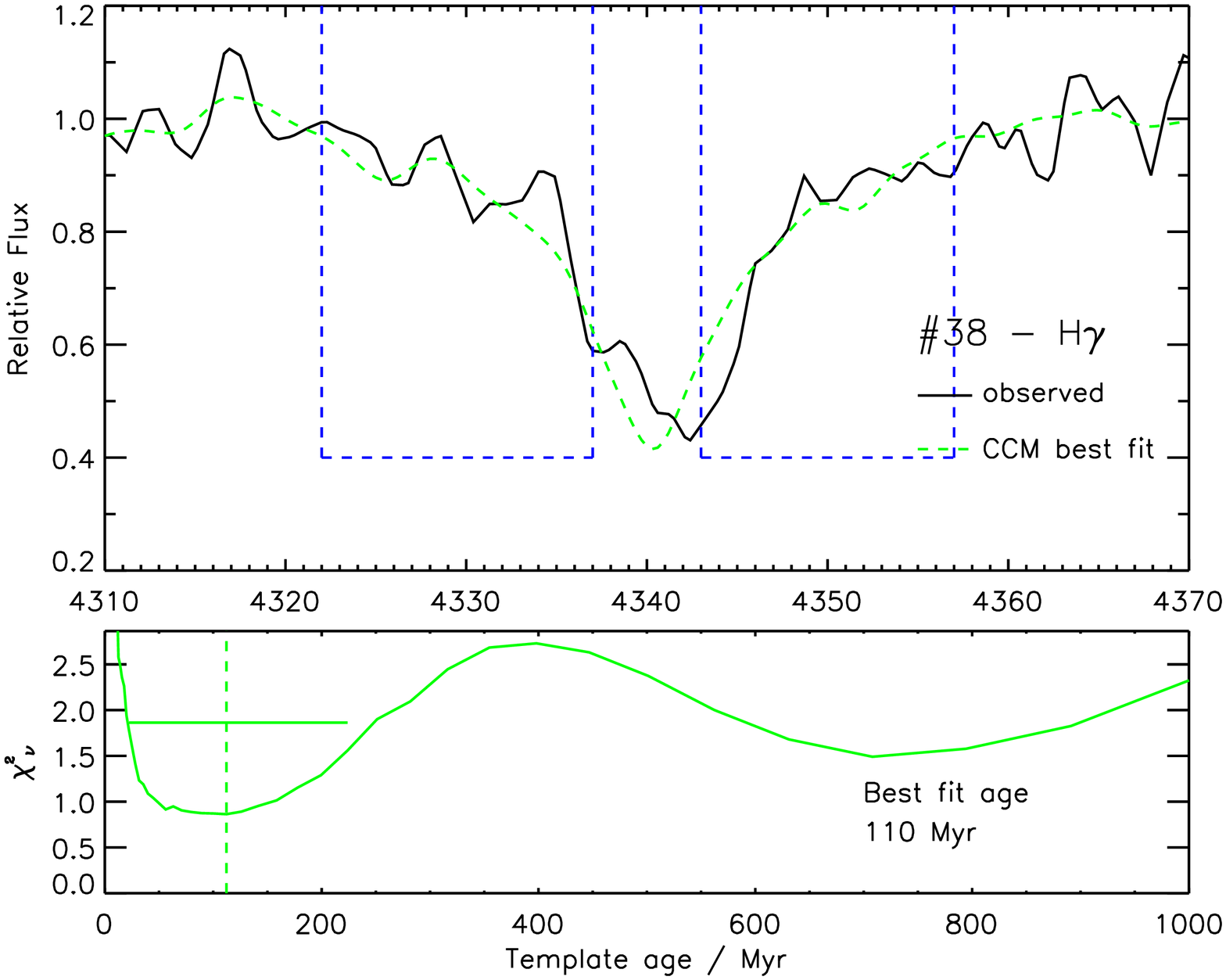}\vspace{5pt}
\includegraphics[width=0.37\textwidth]{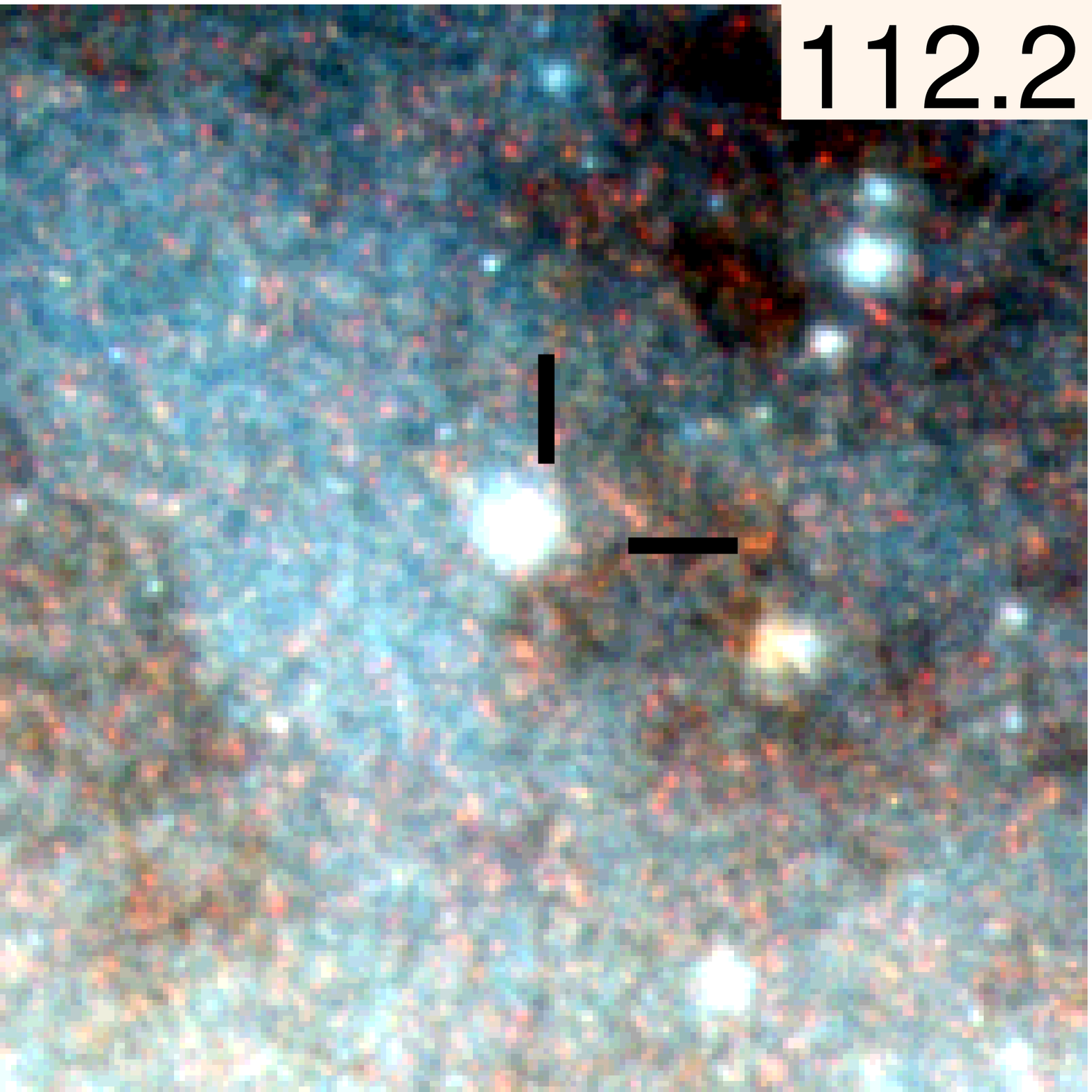}\hspace{5pt}\includegraphics[width=0.46\textwidth]{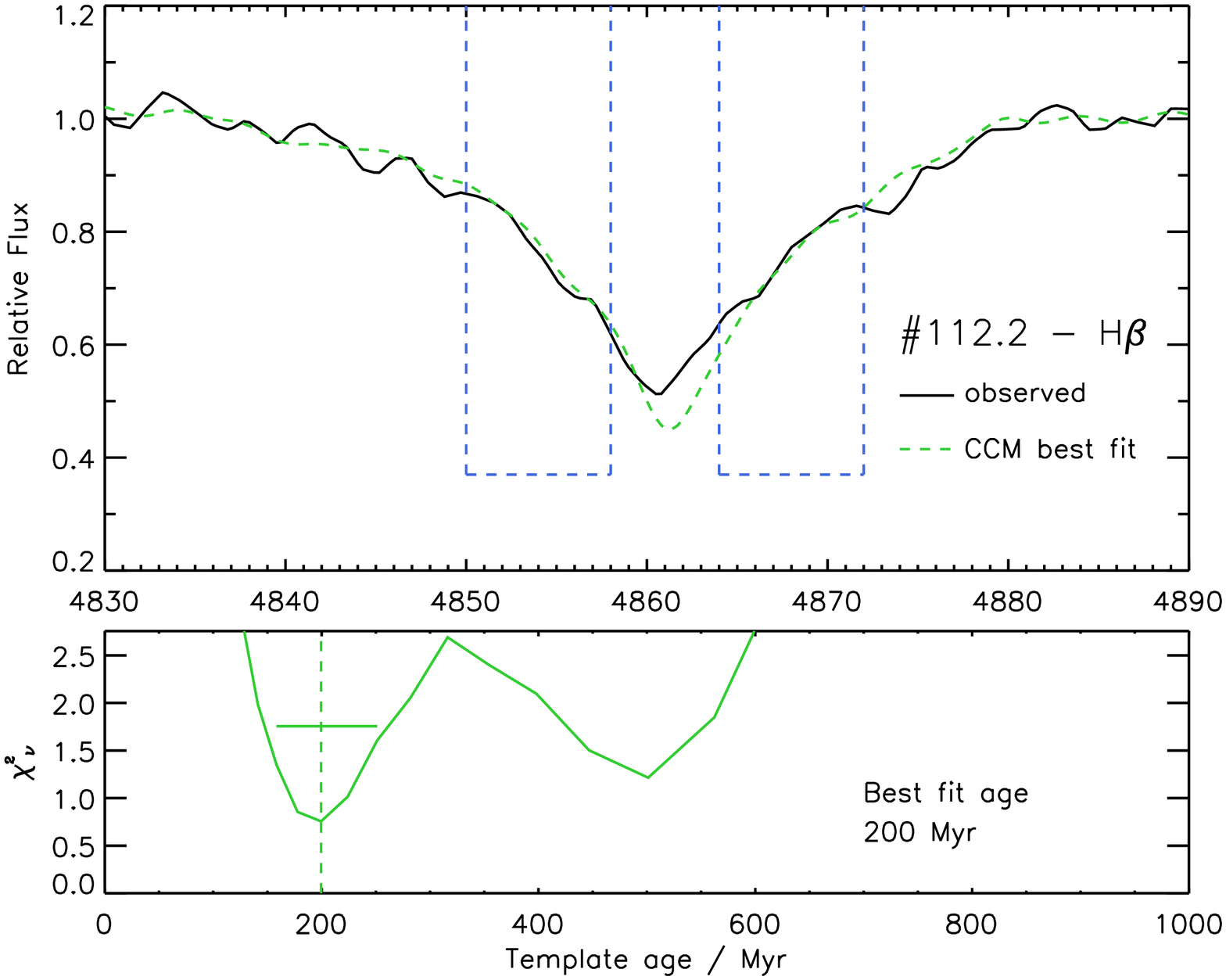}
\caption{(Previous page). Cluster age dating. Each of these three image-spectrum pairs represents a {\it BVI} composite image and part of its optical spectrum, as used in our age-dating method. In the images, East is to the left and North towards the top, with the major axis of the galaxy inclined at $\sim30$~degrees to the horizontal. The spatial scale is a square size of 200~px, or 175~pc. In the spectra, the top panel shows the fit, with the solid line and dashed (green) lines representing the observed spectrum and best fitting model respectively. The dashed blue boxes indicate the spectral regions where the fit takes place. The bottom panel shows the probability distribution of the CCM across age-space, with a vertical line indicating the best fitting age, while the horizontal line indicates the error on the fit ($\chi^2_{min}+1$).\newline
{\bf Top:}~Cluster 19, in the western side of the galaxy disk. This is an example of a low S/N spectrum (${\textup S/N}<20$ at 5000\AA).  
The faint and strong absorbing dust lanes to the west and east of the cluster provide a demonstration of the sub-parsec spatial scale and the amount and strong effect of the galactic dust in M82. 
This cluster has two high probabillity regions in age-space, however, the CCM draws its best fit from the low age trough, placing the age at $\sim40$~Myr.\newline
{\bf Middle:}~Medium S/N ($\sim25$) spectrum of cluster 38. Note that the fit presented here is on the H$\gamma$ line, and the S/N is lower at the blue end of each spectrum. \newline
{\bf Bottom:}~High S/N ($\sim40$) spectrum of cluster 112.2, situated north of region~B. Note that in this case, all age measurements (for both lines) agree on 199~Myr. 
}
\label{plot:age-fits}
\end{center}
\end{figure*}

One immediately noticeable feature in these fits is the existence of two best-fit troughs. While the two fits are statistically equivalent, there is a significant volume of evidence that favours the younger ages. This is based on the comparison to ages independently derived through \textit{UBVI} photometry, as well as a number of considerations of the star formation history of M82 that will be discussed in \S~\ref{sec:age-dist}. The age measurements are shown in Table~\ref{tab:res}, where we have listed the acceptable range as the extent of the preferred best fitting trough.

A final consideration concerns the effect of our choice of model library. We used the GD05 templates to achieve the highest quality fits, given their high spectral and temporal resolution. As a consistency check, we fitted some of the high S/N clusters in our sample with BC03 \citep{bc03} models of similar characteristics (same isochrones, metallicity, and IMF) and obtained very similar results. For instance, cluster~91 was found to have an identical age under GD05 and BC03. Further in support of our choice of templates is the fact that, under BC03, the best fit troughs are represented by very few consecutive models (perhaps three or four), whereas with GD05 the $\chi^2$ probability distribution appears smooth and therefore more reliable.

Having measured the age of each cluster, we can now determine their radial velocities (RVs). We used the IRAF FXCOR routine to cross-correlate the best fitting model with the cluster spectrum. We also manually measured the velocities of the Na\one\ D lines ($\lambda\lambda5890, 5896$) which originate from cold interstellar gas along the line of sight (LOS). We transformed the measured cluster velocities to the heliocentric frame of reference and corrected for the systemic velocity of M82 \citep[200\kms;][]{mckeith93}. The final RV values vary between $0-220$\kms, with associated Na\,I~D velocities in the range $5-200$\kms\ (in absolute value). These measurements are also presented in Table~\ref{tab:res}. We also used the radial velocities to refine our age measurements: we repeated the age fitting process, this time applying a more accurate Doppler-shift according to the derived RVs. This allowed for  a better alignment between spectrum and model, which had been estimated in the first iteration. This second iteration of the age fits did not alter any of the best fitting ages.

We divide our sample into disk clusters and nuclear region knots, where the nuclear region is defined as a projected radius of $\sim200$~pc about the dynamical centre of the galaxy \citep[see][for exact location]{mckeith93}. Note here that we use the terms `cluster', `knot' and `cluster complex' interchangeably for these nuclear region objects; they are probably not individual clusters but cluster complexes (physically associated) or asterisms (chance projections). We find all disk clusters to have ages between $\sim30$ and 270~Myr, with a mean (and median) value of $\sim140$~Myr. For the knots in the nuclear region (slit numbers 67 and 78), these values range from 7 to 30~Myr with a mean of 10~Myr. Unfortunately, it is not possible to obtain accurate age dating in the nuclear region as the fits are dominated by strong emission lines. This restricts the fitting to a spectral range that hardly covers the Balmer absorption lines. The CCM picks very young ages due to the flatness of these spectra; in the case of knots in slit 67, the fit favours the very youngest models in GD05 library, as all models above a few 10~Myr feature noticeably stronger metal lines\footnote{The fit extends to the regions adjacent to the Balmer lines, where weak metal lines are present.}. Further evidence of the young age of these clusters is offered by the strong, embedded emission lines. Such strong features centred at the same wavelength as the absorption can only be associated with the clusters. The clusters in slit 78 display a similar appearance, however, with some more structure (i.~e. they are not as `flat' as the sources in slit 67) thus they are assigned slightly older ages (in the third and fourth decade of Myr). In addition, we find offset emission in cluster 78.1 perhaps indicating that it is not situated deep in the galaxy like the rest of the `nuclear' knots. Finally, our sample also contains two halo objects, which we have dated at $>8$~Gyr; we discuss these candidate old populous clusters (OPC) in \S~\ref{sec:gcs}.

\subsection{Cluster photometry: colours and extinction}\label{sec:col}
We use the photometric measurements to derive cluster colours ($B-V$ and $V-I$) and present these results in Fig.~\ref{plot:colours}. Here we compare the observed cluster colours with GALEV evolutionary models for SSPs \citep{anders03-galev}. {As for our chosen SEDs, the photometric models} use the Padova isochrones, a Salpeter IMF and solar metallicity. 

\begin{figure}
\begin{center}
\plotone{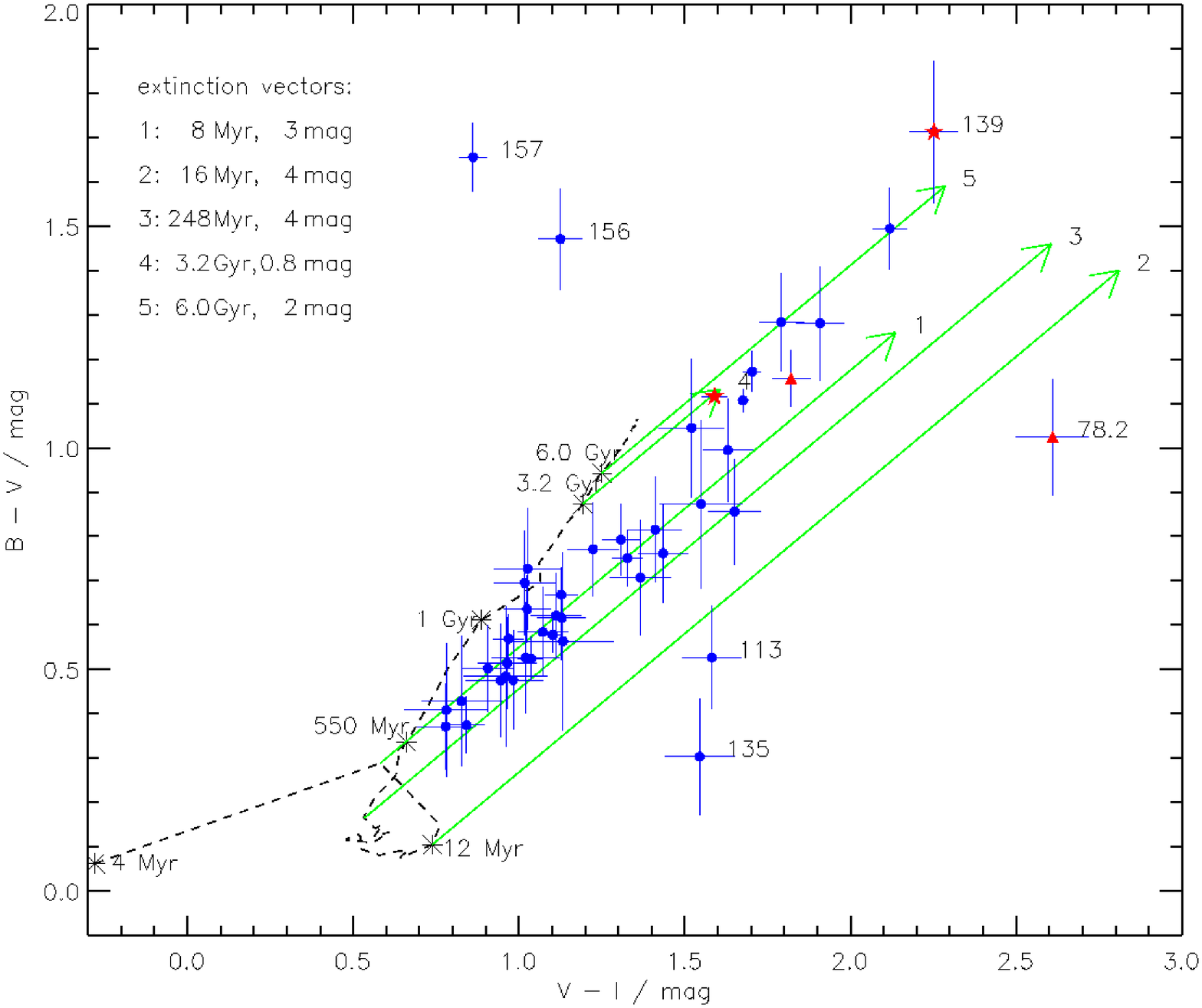}
\caption{$B-V$ vs $V-I$ colour-colour diagram for all disk clusters with available spectroscopy; we have marked off-disk OPC candidates with stars and nuclear knots with triangles.
Foreground-corrected cluster extinctions range between near zero and 2.7~magnitudes ($A_V$) in the disk and $1.4-3.0$~mag in the nuclear region. The concentration of datapoints around the 250~Myr extinction vector reflects the limited age scatter about that age. The two indicated pairs of isolated clusters (nos. 113, 135, 156 and 157) have underestimated fluxes in one or more of the three bands. 
}
\label{plot:colours}
\end{center}
\end{figure}

This plot can be used as a rough diagnostic of age but by no means an accurate one, given  the lack of information in the Balmer jump region and the consequent degeneracy in the $BVI$
 baseline. From this it becomes obvious that there is no clear distinction when age-dating photometrically between a cluster of, e.~g. age 4~Myr and 2.5~Gyr. Given some {\it a priori} knowledge of extinction, only some rough deductions can be made.

More importantly, given a strong prior knowledge of age (which we possess from spectroscopy), colours can be used to estimate the reddening for each cluster. To do this, we used a simple code to transfer each cluster colour along the extinction vector until it is close to the intrinsic (model) colour at the derived spectroscopic age. Assuming a standard Galactic extinction law with $R_V=3.1$ \citep{rieke}, we measured extinctions  in the M82 galaxy disk in the $A_V$ range 0.4--3.2, with a median value of 1.3~mag\footnote{This range excludes cluster 158, a faint cluster with highly uncertain photometry}. The corresponding range of values for the nuclear region is 1.4--3.5 with a  median of 2.2~mag. To derive approximate errors for the $A_V$ values, we have adopted an age range of 50~Myr (a representative value for the error in age measurement) and calculated the corresponding extinctions.

All values are listed in Table~\ref{tab:phot}. Note that the foreground extinction from the Milky Way is $\sim0.5$~mag in the $V$-band, so the lowest values we derive correspond to roughly no extinction from material associated with M82. Interestingly, the measured values do not reflect the large amounts of extinction one might expect to find in this galaxy -- reaching 15 magnitudes, as we mention in the introduction. This is due to our sample, which sought out the brightest possible sources to secure the acquisition of high S/N spectra. As a result, no highly extinguished sources were included.

\subsection{Observations of off-disk clusters}\label{sec:gcs}
Our spectroscopic dataset includes two halo objects located at a projected distance of $\sim400$~pc to the south of the eastern galaxy disk, selected from the initial masterlist as candidate old populous clusters. We fit those objects (nos.~152 and 139) with GD05 models of $\frac{1}{5}\Zsun$ (a value characteristic of old globular clusters) and find best fit ages of~$\gtrsim8$~Gyr. In the case of 152, we also find a peculiar/halo radial velocity as it lags significantly behind the disk population. Unfortunately, the low S/N of cluster 139 does not allow for a certain age fit or a high-confidence radial velocity cross-correlation, therefore only 152 was investigated further (see \S~\ref{sec:gccol}). These clusters have not been observed in the past \citep[they are not included in the list by][who studied a sample of GC candidates around M82 and reviewed the case for GCs in the galaxy]{saito05} and make very interesting targets for further study.

\subsection{Detection of WR features in the nuclear region}\label{sec:wr}
The designed multi-object mask features two long slits placed over the M82 nuclear region, covering a total of seven knots. The precise location of these sources is not easy to pinpoint, given the apparent complexity of the nuclear region. We therefore reserve the possibility that they lie near the surface of the galaxy and are projected onto the nuclear region (this matter is revisited in \S~\ref{sec:discussion:rv}). In addition, the extracted apertures are likely to contain blended light from more than one star cluster, therefore we do not treat these as observations of individual sources. The activity in the nuclear region is key to understanding the starburst history of M82. A recent study by \citet{barker08}, found no evidence of cluster formation over  the last $4-6$~Myr in the nuclear region. However, we find  Wolf-Rayet (WR) features in the spectra of three clusters in the region (maybe cluster complexes; sources 1, 2 and perhaps also 4 within slit 67), which are indicative of such young ages. More specifically, we find evidence of He\two\ $\lambda4686$, C\three\ $\lambda4651$, and possibly C\four\ $\lambda4659$, N\three\ $\lambda4634-4640$ and N\five\ $\lambda4603-4619$; all these features form part of the `blue WR bump' at $4600-4700$~\AA\ \citep{sidoli06,bibby08}. In Fig.~\ref{plot:id67} we provide a finding chart for slit No.~67 and we present the spectra of apertures 1 and 2, indicating the locations of the proposed WR features. This detection needs to be confirmed with higher S/N in this spectral region, and if proven to be true, it will classify M82 as a WR galaxy. 
 
\begin{figure*}
\begin{center}
\includegraphics[width=0.42\textwidth]{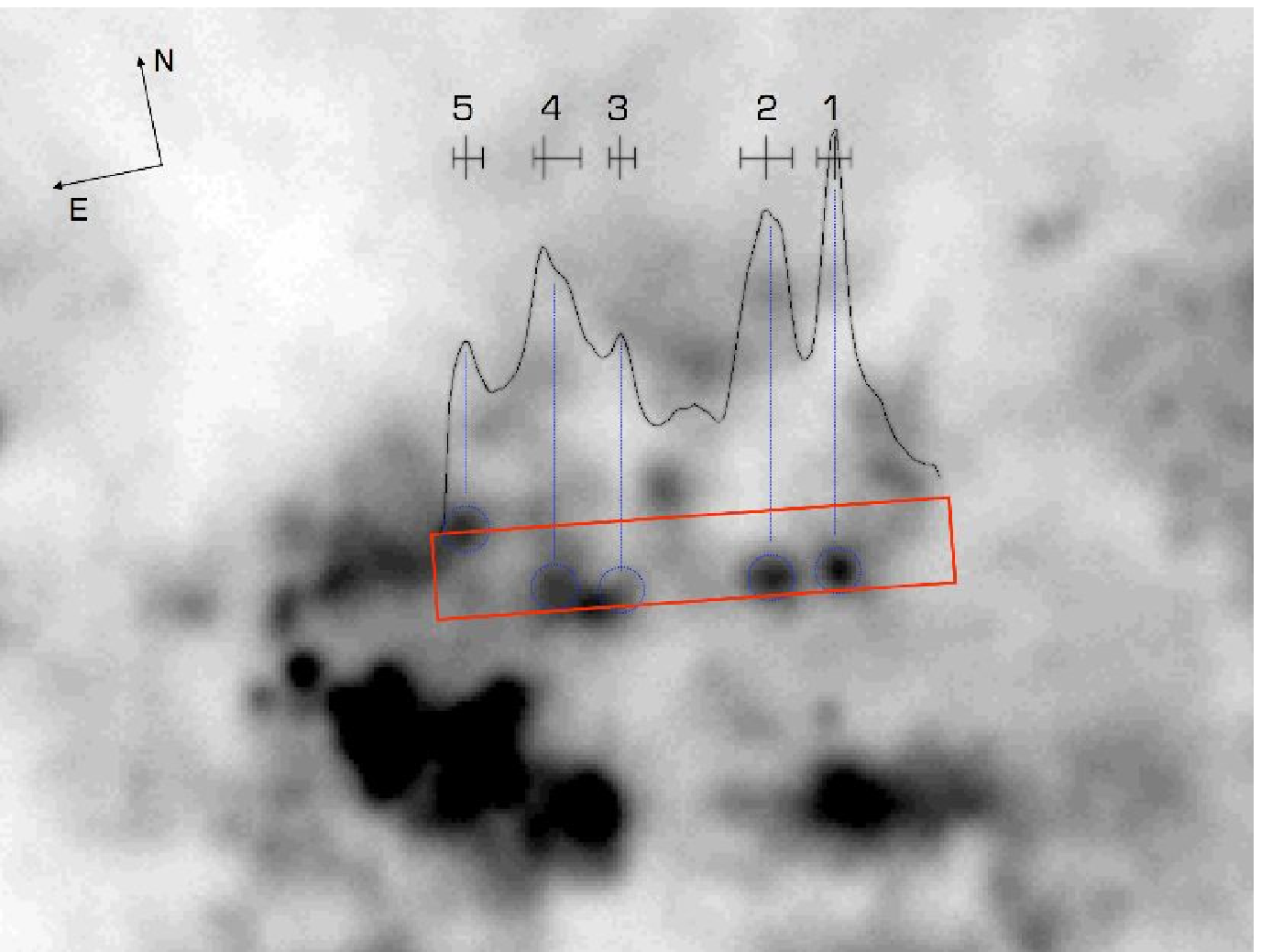}\hspace{5pt}\includegraphics[width=0.53\textwidth]{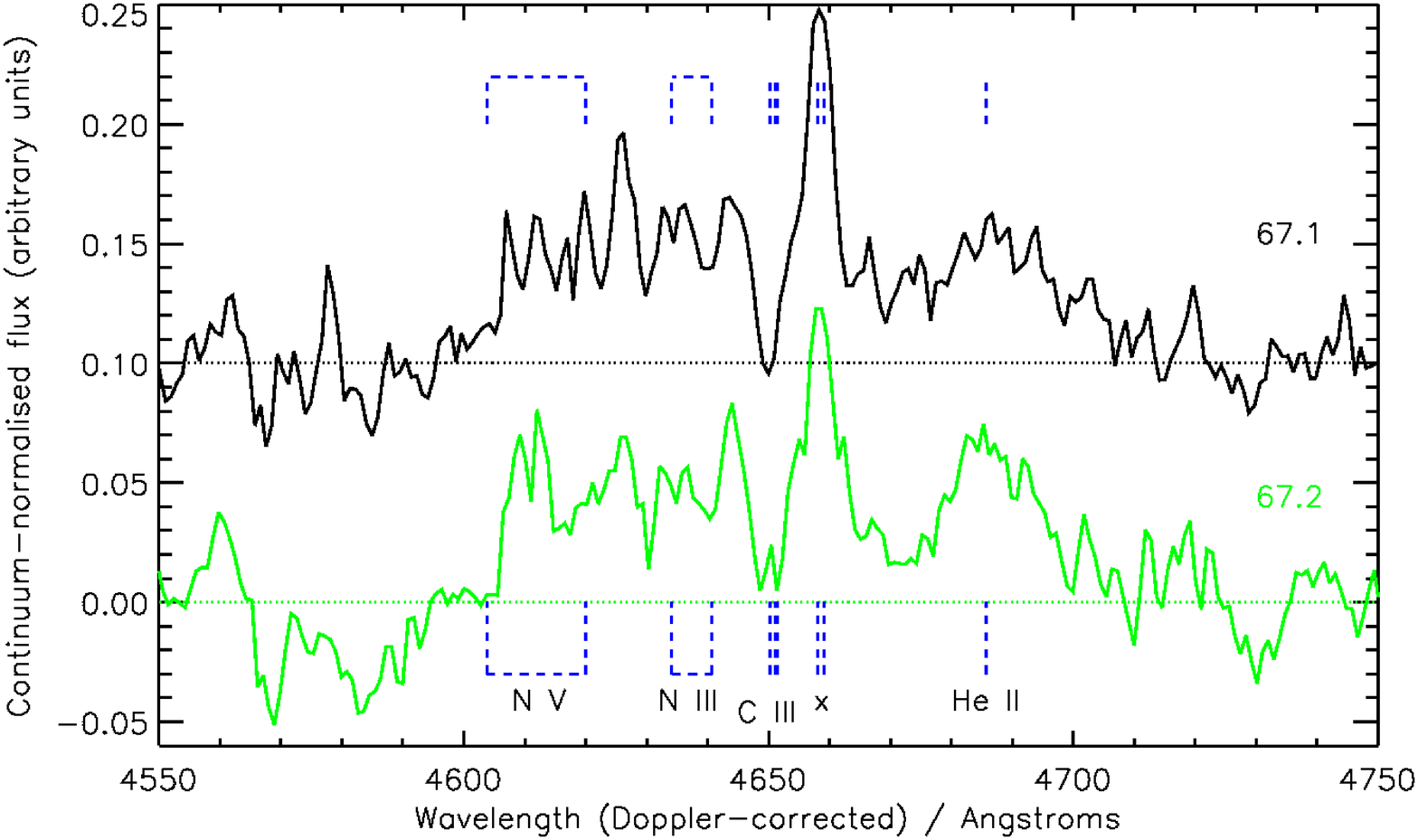}
\caption{
GMOS {\it g}-band imaging of the nuclear region of M82. {\bf Left}: Positioning of spectrum extraction apertures in slit no.~67. The red box represents the exact positioning of slit~67 (tilted by $4.1^\circ$ with respect to the major axis). We have overplotted the integrated light contained within the slit to identify the extracted objects, indicated by the dotted blue circles. {\bf Right}: Extracted spectra for apertures 1~(black) and 2~(green/grey) in slit 67 showing the features, that usually form part of the `blue WR bump' ($\sim4600-4700$\AA). We have normalised the spectra using two continuum windows just outside the plotted range, to show the WR bump. Note that, although we extracted these spectra using narrow apertures to avoid nebular contamination, we cannot reject the possibility that the line marked `$\times$' is in fact nebular [Fe\three] (as opposed to the C\four\ WR feature).}
\label{plot:id67}
\end{center}
\end{figure*}
 
We stress at this point that the exact source of these emission lines is uncertain. They could originate from (or be enhanced by) scattered light from the nuclear region, rather than the observed knots. In order to avoid contamination by nebular features in our analysis, we have extracted these spectra in narrow apertures. Despite having taken that precaution, we reserve the possibility that the feature at $\lambda4658$ is in fact nebular [Fe\three].  In essence it is not the source of this emission that we report here, but simply its existence. Even though our data are not suitable for an in-depth analysis of the detected lines, they provide evidence that the nuclear starburst is still active. WR features only appear in the youngest stars ($2-6$~Myr), and are a reliable age-dating tool for young star clusters \citep{sidoli06,bastian06complex}. More evidence for an ongoing nuclear starburst was offered by \citet{ljs06}, who found an age of 6.4~Myr for cluster A1 in the nuclear region. The high mass of this cluster ($\sim10^6$~\Msun) implies the existence of a fully sampled cluster mass function at that era, and therefore the contemporary formation of a large number of clusters -- such a massive cluster represents but the `tip of the iceberg' in terms of the underlying distribution. Hence we confirm that star/cluster formation is currently continuing in the nuclear region of M82.

Additionally, it was proposed by \citet{barker08} that, while stars are forming in the nuclear region of M82, star and cluster formation have become dissociated in this region. Our detection of young clusters in the same region studied by \citeauthor{barker08} indicates that star and cluster formation is still ongoing. Furthermore, \citet{bastian08sfr} showed that cluster A1 fits the relation between the brightest cluster in a population and the SFR of the host galaxy (a relation which extends from quiescent to extreme starburst galaxies).  This suggests that star and cluster formation is proceeding normally in the nuclear region -- subject to our interpretation that these clusters reside in the nucleus (see \S~\ref {sec:discussion:rv} for further discussion).

\begin{table*}
\caption{Spectroscopically derived ages and radial velocities for all clusters in the sample. 
}
\label{tab:res}
\begin{center}\scriptsize
\begin{tabular}{%
p{1.3cm}
p{0.6cm}
p{0.6cm}
p{0.6cm}
p{0.8cm}
p{1cm}
l
p{1cm}
r
p{0.4cm}
r
p{0.5cm}
r}
\tableline
\tableline
ID (alt ID)	&	S/N - H$\gamma$	&	S/N - H$\beta$	&	$\tau_{\rm min}$ & $\tau_{\rm best}$ & $\tau_{\rm max}$	&	~~~~Mass & $\pm$	&	$u_R\tablenotemark{a}$	&	$\pm$	&	$u_R^{Na\,I~D}$	&	$\pm$	&	$d_{GC}$	\\
	&					&				&	& Myr & 	&	\tmult{\Msun}	&	\multicolumn{2}{c}{\kms}	&	\multicolumn{2}{c}{\kms}	&	pc		\\
\tableline
%
       7 &	18 &	  23 &        60 &	 120 &       200 &   3.6e+04 &   1.7e+04 &	-159 &       141 &	  44 &        10 &	2520 \\ 
       8 &	13 &	  22 &        30 &	 100 &       160 &   2.8e+04 &   1.4e+04 &	-150 &        14 &	 -24 &        10 &	2385 \\ 
      14 &	 5 &	  17 &        30 &	 160 &       890 &   8.9e+04 &   2.8e+04 &	-103 &        23 &	  -9 &         1 &	1909 \\ 
      17 &	 9 &	  11 &        90 &	 280 &       710 &   1.1e+05 &   4.9e+04 &	-182 &         0 &	  -6 &         2 &	1763 \\ 
      18 &	23 &	  30 &        60 &	 120 &       180 &   9.5e+04 &   4.0e+04 &	-127 &        15 &	 -37 &         7 &	1735 \\ 
      19 &	 7 &	  18 &         7 &	  40 &       130 &   2.6e+04 &   1.2e+04 &	-125 &        21 &	 -58 &        12 &	1597 \\ 
      20 &	14 &	  21 &        10 &	 140 &       130 &   2.8e+04 &   1.5e+04 &	-115 &        24 &	 -75 &         9 &	1525 \\ 
      21 &	 9 &	  18 &        60 &	 200 &       350 &   7.1e+04 &   3.2e+04 &	 -64 &        27 &	  -7 &         6 &	1430 \\ 
      28 &	17 &	  30 &        90 &	 150 &       280 &   9.2e+04 &   3.2e+04 &	-105 &        30 &	 -27 &         5 &	1221 \\ 
      29 &	17 &	  26 &        30 &	  80 &       130 &   1.4e+05 &   4.4e+04 &	-106 &        15 &	 -50 &         6 &	1175 \\ 
      33 &	25 &	  27 &        80 &	 140 &       140 &   3.0e+05 &   8.0e+04 &	-103 &        30 &	 -46 &        13 &	1036 \\ 
      34 &	11 &	  27 &        30 &	  90 &       200 &   2.9e+05 &   7.9e+04 &	 -94 &        29 &	 -36 &        10 &	1029 \\ 
      38 &	19 &	  27 &        90 &	 130 &       200 &   8.5e+04 &   3.2e+04 &	-132 &        18 &	   5 &         9 &	 814 \\ 
    43.1 &       8 &	  27 &	      90 &	 160 &	     220 &   1.4e+06 &	 3.4e+05 &      -212 &	       0 &	  20 &	       2 &	 750 \\ 
    43.2 &      15 &	  23 &	     100 &	 130 &	     320 &   1.4e+05 &	 6.0e+04 &	 -34 &	      42 &	  15 &	       1 &	 640 \\ 
      51 (F\tablenotemark{b}) &	68 &  41 &  40 &  70 &        80 &   4.5e+06 &   1.3e+06 &	-157 &       100 &	  32 &         6 &	 612 \\ 
      58 &	17 &	  22 &        10 &	  30 &        70 &   4.6e+04 &   1.9e+04 &	  -9 &        36 &	  64 &        16 &	 445 \\ 
    67.1 &	73 &	  76 &         4 &	   7 &        10 &   7.2e+04 &   2.0e+04 &	-153 &       120 &	   9 &         5 &	 297 \\ 
    67.2 &	42 &	  75 &         4 &	   7 &        10 &   8.9e+03 &   2.7e+03 &	 -67 &        50 &	  20 &         2 &	 264 \\ 
    67.4 &	26 &	  57 &         4 &	   7 &        10 &   7.7e+03 &   2.4e+03 &	 -15 &        49 &	  22 &         7 &	 151 \\ 
    78.1 &	24 &	  26 &        20 &	  30 &        40 &   1.6e+04 &   6.6e+03 &	   7 &        49 &	  31 &        24 &	 102 \\ 
    78.2 &	57 &	  57 &        10 &	  20 &        20 &   3.0e+05 &   7.2e+04 &	  28 &        61 &	  35 &         6 &	  56 \\ 
      91 (H\tablenotemark{c}) &	41 &  44 & 180 & 200 &       200 &   2.1e+06 &   6.0e+05 &	  70 &        36 &	  32 &         6 &	-156 \\ 
      97 &	26 &	  35 &        90 &	 160 &       200 &   2.7e+05 &   1.3e+05 &	 114 &        25 &	  64 &        16 &	-274 \\ 
      98 &	42 &	  33 &       130 &	 160 &       200 &   9.3e+04 &   5.9e+04 &	  64 &        14 &	 101 &         8 &	-289 \\ 
   103.1 &	21 &	  23 &        40 &	  80 &       110 &   1.5e+04 &   1.0e+04 &	  90 &         7 &	  57 &         9 &	-338 \\ 
   103.2 &	29 &	  30 &        40 &	  80 &       110 &   2.0e+05 &   7.0e+04 &	  88 &        26 &	  55 &        39 &	-397 \\ 
     108 &	 8 &	  20 &        90 &	 160 &       220 &   2.4e+05 &   7.6e+04 &	 221 &        19 &	  38 &        51 &	-525 \\ 
   112.2 &	36 &	  34 &       160 &	 200 &       250 &   5.8e+04 &   3.1e+04 &	 103 &        11 &	  96 &        11 &	-614 \\ 
     113 &	31 &	  26 &        60 &	 110 &       140 &   1.6e+05 &   5.4e+04 &	 127 &        28 &	 131 &         8 &	-660 \\ 
     125 &	10 &	  21 &        60 &	 140 &       220 &   2.6e+05 &   1.1e+05 &	 176 &        85 &	  26 &        24 &	-847 \\ 
     126 &	35 &	  31 &       140 &	 210 &       350 &   2.9e+05 &   1.4e+05 &	 217 &        20 &	 113 &         4 &	-862 \\ 
     131 &	32 &	  40 &        60 &	  80 &       110 &   3.8e+05 &   1.6e+05 &	  82 &        17 &	 106 &        12 &	-965 \\ 
     135 &	14 &	  22 &        80 &	 140 &       220 &   5.7e+04 &   2.1e+04 &	  82 &        12 &	  77 &        12 &     -1072 \\ 
   138.1 &	11 &	  16 &        20 &	  50 &       130 &   9.3e+03 &   3.7e+03 &	  59 &        19 &	  98 &        15 &     -1221 \\ 
     139 &	 4 &	  14 &       $-$ &	8000\tablenotemark{d} & $-$ &   4.5e+06 &   1.4e+06 &	-212 &         0 &	  21 &         6 &     -1262 \\ 
     140 &	16 &	  17 &        30 &	  60 &       110 &   7.5e+04 &   2.4e+04 &	 119 &         7 &	 108 &        19 &     -1351 \\ 
     142 &	17 &	  34 &       200 &	 220 &       350 &   2.4e+05 &   8.5e+04 &	 146 &        38 &	  85 &         1 &     -1561 \\ 
     143 &	30 &	  32 &       200 &	 240 &       500 &   8.5e+04 &   3.7e+04 &	 147 &        50 &	  93 &         1 &     -1569 \\ 
     145 &	13 &	  18 &        30 &	  90 &       160 &   1.4e+04 &   6.9e+03 &	 115 &        19 &	  94 &         1 &     -1686 \\ 
     147 &	12 &	  22 &        30 &	  70 &       110 &   4.6e+04 &   1.6e+04 &	 146 &        16 &	  21 &        53 &     -1755 \\ 
     152 &	29 &	  29 &       $-$ &	8000\tablenotemark{e} & $-$ &   4.1e+06 &   2.3e+06 &	  62 &        18 &	  34 &         6 &     -1973 \\ 
     154 &	11 &	  22 &       100 &	 180 &       350 &   4.0e+04 &   2.8e+04 &	 160 &        17 &	 105 &        17 &     -2037 \\ 
     155 &	12 &	  18 &        60 &	 210 &       790 &   1.9e+04 &   1.2e+04 &	 149 &        10 &	  85 &         3 &     -2108 \\ 
     156 &	11 &	  26 &        60 &	 130 &       180 &   1.7e+05 &   5.6e+04 &	 149 &        15 &	  51 &         5 &     -2195 \\ 
     157 &	48 &	  47 &       220 &	 270 &       320 &   5.8e+05 &   2.0e+05 &	 115 &        25 &	  92 &         5 &     -2188 \\ 
     158 &	 9 &	  20 &       110 &	 160 &       280 &   1.2e+04 &   1.0e+04 &	 145 &        27 &	 108 &        24 &     -2313 \\ 
     159 &	15 &	  26 &       100 &	 170 &       350 &   4.0e+04 &   1.8e+04 &	 132 &        12 &	  95 &         3 &     -2628 \\ 
     160 &	12 &	  21 &        60 &	 140 &       220 &   1.6e+04 &   9.9e+03 &	 121 &        28 &	 101 &         0 &     -2648 \\ 

\tableline
\end{tabular}
\end{center}
\tablenotetext{a}{The quoted velocities are heliocentric and corrected for the systemic velocity of M82.}
\tablenotetext{b,c}{Following the OM78 notation}
\tablenotetext{d,e}{See \S~\ref{sec:ages-rvs}, \ref{sec:gccol} for discussions on the age and nature of these sources.}
\end{table*}

\section{Discussion}
In this section we discuss the spectroscopic (age and radial velocity) and photometric (colours and reddening) measurements for our sample of 49 disk, halo and nuclear star clusters in M82, and the implications of these results.

\subsection{Colours and reddening}
Cluster colours are presented in Fig.~\ref{plot:colours}; this plot works well in confirming the timescale for the starburst, with many datapoints distributed closely about the 250~Myr extinction vector. In addition, the nuclear region clusters are intercepted by younger age vectors, and the OPC candidates by vectors in accord with their spectroscopic ages. There are two isolated pairs of clusters (nos. 113 and 135, 156 and 157) that do not agree with the spectroscopic ages; of those, cluster 113 appears to have an underestimated flux in blue light, which we interpret as the result of differential extinction across the face of the cluster \citep[this will shift the datapoint redwards in blue bands preferentially; see][and Paper I for more discussion of such cases]{bastian07m82f}.  Cluster 135 appears diffuse in the ACS imaging, and could therefore have disproportionately underestimated fluxes. Cluster 156 has an underestimated B-band flux, as it resides behind a thin, filamentary dust lane; cluster 157 lies near the edge of the optically visible disk, at the boundary between the bright, blue disk and  wide, dark band to the north of region~B, and is therefore differentially extinguished. 

\begin{figure}
\begin{center}
\plotone{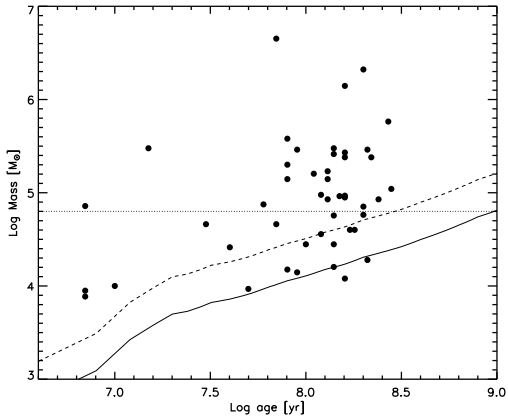}
\caption{Cluster masses plotted against age (the age-mass diagram) for the entire spectroscopic sample presented in the current work. The solid line represents the detection limit at $m_V=21$ (corresponding to $M_V\sim-7$ at the distance of M82) translated into mass. The dynamical nature of this limit is due to the evolutionary fading of star clusters and therefore the variation of the mass to light ratio (the line has been derived using GALEV models \citep{anders04}). We have also shown the result of one magnitude of extinction as a dashed line (the same intrinsic limit coupled with one magnitude of extinction). The limit seems to accommodate all the data-points comfortably within the errors (not shown; of the order of 0.3~dex both horizontally and vertically); the dotted horizontal line shows a mass limit at $\sim6\times10^4$~\Msun, which is the mass to which our sample is predicted to be complete to an age of 1~Gyr; this is expressed as the point of intersection between the mass cut and the detection limit. As a large fraction of our sample lies above that cut-off, we deduce that the non-detection of clusters older than $\sim300$~Myr is does not result from incompleteness), but reflects the intrinsic age distribution of the M82 starburst clusters.
}
\label{fig:completeness}
\end{center}
\end{figure}

We present cluster reddenings in the bottom panel of Fig.~\ref{plot:rot-curve}, where in this case the values have been corrected for foreground galactic extinction \citep[$A_V=0.53$;][]{schlegel98}. The values presented elsewhere in the paper do not incorporate such a correction.

\subsection{Cluster Masses}
Having obtained accurate photometry (magnitudes and extinctions) and ages for each star cluster in our spectroscopic sample, we derived their photometric masses (see Table~\ref{tab:res} for results). We find average and median values of $2.5\times10^5$ and $7.2\times10^4$~\Msun. According to our photometric mass estimates, the most massive individual clusters in our sample are M82-F and H (nos.~51 and 91 on our list) and no.~152 (the candidate OPC), with values of $4.5\times10^6$, $2.1\times10^6$ and $2.1\times10^6$~\Msun. The total baryonic mass contained in the observed clusters sums to $1.1\times10^7$~\Msun.

Based on these mass measurements, we can derive an estimate for the total mass in clusters for the entire galaxy. We generated an array of simple stochastic populations of star clusters, governed by a power law mass function of index $-2$, with a minimum mass of $50$~\Msun\ and no upper mass limit. As we know of five clusters in M82 with mass above $10^6$~\Msun, we then varied the number of clusters in the population, $N$, until the average (over 100 runs) number of clusters with mass greater than $10^6$~\Msun\ is 5. From this we find a total cluster mass for M82 of $\sim10^8$~\Msun, the mass of  a stochastic cluster population with $N=10^5$. This is an approximate lower limit, as there may exist more star clusters in M82 with mass greater than $10^6$~\Msun\ that have not been observed due to extinction.

We repeated this exercise analytically, by calculating the missing mass according to a power law mass function of index $-2$. Given five clusters with mass of the order of $10^6$, we integrate the mass function and extrapolate a mass of $\sim6\times10^7$~\Msun~for the observable population. Taking this consideration further, and assuming that: a) all the clusters we have studied formed during the most recent starburst; and b) that during the starburst, 10\% of stars formed in clusters that are still bound in the currently observable era, then we estimate the mass of the young field star population at $\sim5\times10^8$~\Msun. This value is an underestimate, as it was calculated based on the observable cluster population. In addition, recent studies by \citet{larsen09} and \citet{gieles-cimf} find that the underlying cluster mass function is likely to be a Schechter function, with a limiting mass of $M_*\sim10^6$~\Msun. This is another factor that may lead to a false estimate of this total mass, as we have integrated over a power law with no imposed maximum mass. While this value represents a lower limit, we note that the survival of a significant number of massive clusters to $\sim150$~ Myr argues against a prolonged era of  mass-independent cluster disruption.

\subsection{Age distribution}\label{sec:age-dist}
Using the spectroscopically derived ages, we have plotted the age distribution of all clusters in the sample in Fig.~\ref{fig:age-dist}, where the solid and dashed lines represent the disk and nuclear region cluster populations respectively. We chose to bin the data in 50~Myr bins to reflect the errors in some of the measured ages; {also, the first bin has half the size of the rest, to mirror the lack of disk clusters younger than $\sim40$~Myr}. We note that the overall shape of the histogram is maintained when a smaller bin size is applied, however the low number of clusters introduces uncertainties. {The plotted error bars account for Poissonian sampling and go as $\sqrt{N}$ for each bin}. As the spectroscopic data sample the entire galaxy disk and nucleus, this can be treated as a rough indicator of the M82 star formation history (SFH), where the number of clusters formed reflects the mean star formation rate (SFR) at that epoch. Naturally, the numbers cannot be treated as correctly `to scale' as our dataset is not complete. 

\begin{figure}
\begin{center}
\plotone{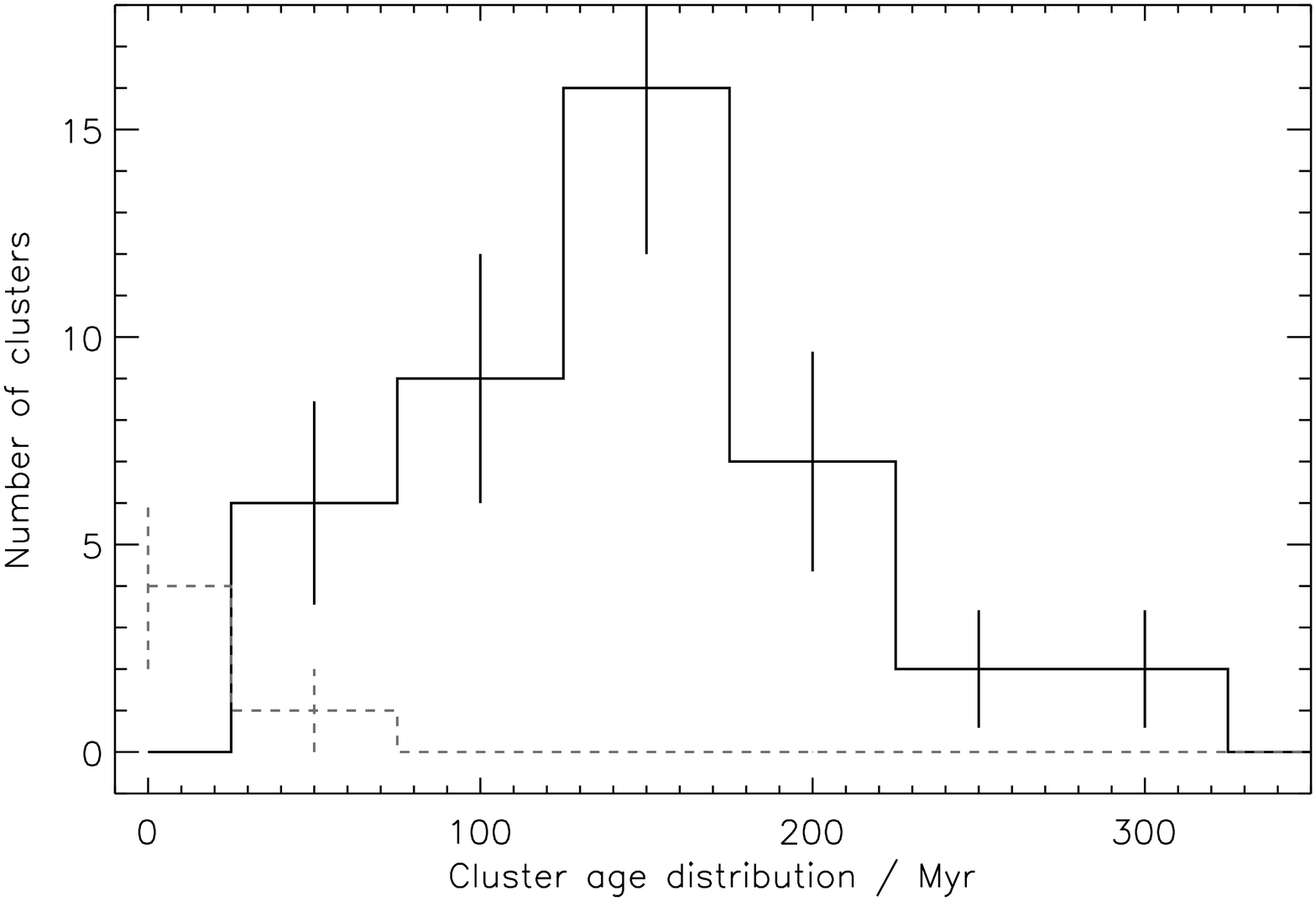}
\plotone{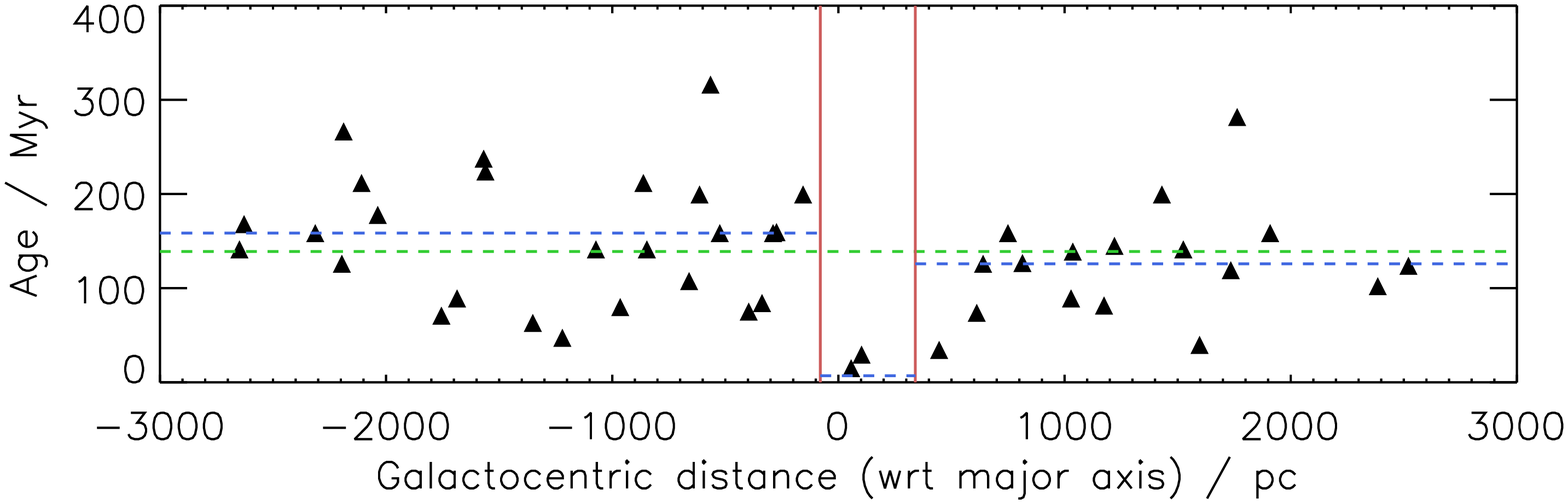}
\caption{{\bf Top:}~Distribution of cluster ages over the past 300~Myr (this range contains all observed clusters). This is a rough representation of the cluster formation history of M82, as the sample is incomplete. The solid and dashed lines traces cluster formation in the disk and nucleus respectively. As the spectroscopic sample spans the entire galaxy disk and nucleus, the number of clusters formed can be used as a diagnostic of the star formation rate, in a relative scale. The burst is placed at 220~Myr, and its peak SF epoch at $\sim150$~Myr, after which the SFR gradually declines. After reaching an apparent post-starburst minimum at $\sim50$~Myr, the disk starburst appears to halt, while the nuclear burst leads the SF in the galaxy.\newline
{\bf Bottom:} Spatial distribution of ages for the spectroscopic cluster sample. The vertical lines delimit the nuclear region and the horizontal lines mark the median age for the region over which they span (i.e. western disk, eastern disk, nucleus, and entire sample in green). This shows no concentration of a certain age group in any region of the disk, indicating a global starburst event lasting the entire range of measured ages. 
}
\label{fig:age-dist}
\end{center}
\end{figure}

The histogram confirms some of the `milestones' we have discussed in our recent work on the M82 cluster population \citep[][and Paper~I]{smith07regb, bastian07m82f}. First, by placing the burst between $200-250$~Myr, it reflects the timescale for the last encounter with M81 at 220~Myr \citep{yun99}, the event that most likely triggered the disk starburst and stripped away part of the disk. Note that this marks the onset of the disk starburst, not the period of maximum formation that follows closely. This sets another observational constraint on the age of the disk starburst \citep[this was discussed in Paper~I and ][for region~B]{smith07regb}. Finally, the first bin reflects the ongoing nuclear starburst. 

In addition, the age distribution peak for the full disk sample (that we place roughly at 150~Myr) is consistent with the 156~Myr distribution peak for region~B, as derived in~\citet{smith07regb} based on a sample of 35 UBVI age dated clusters. This supports our interpretation of region~B as a regular part of the galaxy disk \citep[][Paper~I]{smith07regb}.  

In \S~\ref{sec:ages-rvs} we discussed the possibility of a degeneracy in our cluster age dating technique. This is expressed through the existence of two best-fit troughs (occurring in the majority of cases), roughly placed at $200\pm100$~Myr and $600\pm100$~Myr. This degeneracy in line profile arises because of  the evolution of the Balmer line profile both in strength and width. These line properties fluctuate in the first few hundred Myr, before commencing a more or less monotonic evolution. More specifically, stellar Balmer lines gradually broaden until the age of $\sim300$~Myr and become narrower once again at $\sim600$~Myr. This makes it difficult to distinguish between spectra either side of this 300~Myr mark using a statistical method. One deduction can therefore be made with certainty: that the clusters we observed are all younger than $\sim600$~Myr. This adds to the findings of Paper~I, where we concluded that the population of region M82-B is not as old as previously believed \citep[in excess of 1~Gyr, as presented in][and consequent works]{RdG03a}. This still does not exclude the possibility of a broader age distribution for M82, possibly stretching to $\sim600$~Myr.

In order to discern the better fit let us consider the existing evidence. 
First, the aforementioned study presented in \citet{smith07regb} found the (independently derived) age distribution to be extremely similar to the one presented in Fig.~\ref{fig:age-dist}: virtual inactivity prior to the burst at $\sim250$~Myr and a gradual quenching in more recent times. In addition, this distribution is centred at $\sim150$~Myr and finds no clusters older than $\sim300$~Myr.
Second, adopting the older age range in these cases results in an extended distribution with a gap between $\sim300-500$~Myr; this cannot be explained without employing complex scenaria such as multiple bursts\footnote{This, in turn, cannot be explained given the current observational evidence, as the burst was most probably caused by the last (and probably only) passage about M81}.
In addition, recent work by \citet{mcquinn09} suggests that a starburst should last no longer than $\sim400$~Myr, therefore advocating against an extended distribution for the starburst cluster population. 
Finally, a note on the method itself: the two-minimum structure does not always exist in both the H$\beta$ and H$\gamma$ fits: For instance, cluster 112.2 (presented in Fig.~\ref{plot:age-fits}) only shows a double minimum in the H$\beta$ fit. In the H$\gamma$ fit there is only one trough, describing the same age range as the H$\beta$ fit.

In light of this evidence, we adopt the younger ages for the all clusters where this degeneracy is present. 

Incompleteness is an issue that needs to be addressed in any study of star clusters in extra-galactic environments; an analysis that is applied to a fraction of a population can lead to misinterpretations \citep[as treated in][]{smith07regb,gieles07b}. In this case, the lack of clusters with age greater than $\sim300$~Myr could be the effect of such an observational bias. The completeness of our sample was estimated by inspecting the age-mass diagram of the clusters and by adopting a conservative detection limit of $m_V=21$ (see Fig.~\ref{fig:completeness}). We find that if a significant population of clusters with ages between 300~Myr and 1~Gyr and masses above $\sim10^4$~\Msun~existed, it would have been included in our sample (assuming that they formed with the same intensity as their younger counterparts).  We therefore confirm that the decrease in the number of clusters with ages greater than $\sim$~300 Myr (see Fig.~\ref{fig:completeness}) is real and not a result of incompleteness. In addition, the lack of older clusters in the photometric (\textit{UBVI}) sample presented in \citet[][where the limiting magnitude was fainter]{smith07regb} also argues for a statistically viable sample.

Thus we conclude that the full disk of M82 (mostly likely including the nucleus) entered an era of increased star/cluster formation rate no more than $\sim300$~Myr ago, which continued in the disk until at least $\tau\sim50$~Myr, and is  ongoing in the nucleus. Furthermore, no area in the galaxy, other than possibly the nuclear region, appears to have a cluster formation history distinct to the rest of the disk.  Unfortunately, we cannot derive the absolute cluster formation history for M82 from this dataset due to sample selection and detection limits \citep[e.~g.][]{bastian05a} and uncertain knowledge of the cluster disruption timescale for this galaxy \citep[e.~g.][]{gieles05,lamers05}. The bottom panel of Fig.~\ref{fig:age-dist} combines the age and spatial distribution of clusters and clearly shows the scatter of all disk cluster ages about 150~Myr, and the very young ages of all nuclear clusters. It also shows the lack of young clusters at large galacto-centric distance: beyond $\sim2$~kpc, no clusters lie significantly below the distribution mean. This is unlikely to be a selection effect, as the brightest clusters were chosen for this programme, and these are most likely to be the youngest ones.

\subsection{On the star formation history of M82}
We will now use the presented age measurements to investigate the star formation history of M82. Although the sample is incomplete, star cluster ages can still be used as rough indicators of the CFR of the galaxy. 

The observational evidence for SF activity in the disk is as follows: there appears to be little or no recent (in the last Gyr or so) SF, prior to the onset of the starburst. The starburst itself is observable in the increased cluster formation at $\tau\sim200$~Myr. The overall CFR in the disk appears to have dropped steadily since the peak starburst epoch at $\sim150$~Myr, reaching an apparent hiatus $\sim50$~Myr ago. However, given the errors inherent in this analysis (mainly due to incompleteness), we cannot firmly establish whether the CFR did indeed decrease or if it remained stable. Furthermore, our sample contains no evidence for the existence of young clusters in the outer regions of the galaxy disk, with younger disk clusters being situated closer to the nucleus. This deduction can be assumed to be true with some degree of confidence, as a young cluster forming region would shine bright in the outer regions, where the extinction is below average. Also, our observations were planned in such a way as to include the brightest (and therefore youngest) available sources; the ages of these clusters place their formation in the first $\sim100$~Myr after the onset of the burst. These observations pose questions about the mechanisms that maintained the M82 starburst over the past $\sim250$~Myr. 

The nuclear burst appears at first glance to offer a simpler picture, with zero activity for most of the burst, and intense cluster formation in the current era. However, incompleteness effects are very grave in the nucleus, the result of both technical/observational and physical factors: first, the nuclear region spreads over a mere $\sim200$~pc, where only two slits could be accommodated. Second, older clusters will be outshone by the very youngest clusters and cluster-forming clumps. In effect, the nucleus may have been producing clusters at a steady rate over the course of the disk starburst, as material was funnelled into the regions around the bar.  Furthermore, the precise location of the observed knots is not simple to pinpoint in this optical dataset. We discuss this further in the following section.

Having considered this body of evidence, we propose the following scenario for the SFH of M82: following a prolonged era of low activity in the disk, the SFR increased rapidly, probably triggered by the encounter with M81, some $\sim220$~Myr ago. As the starburst spread across the disk, the SFR increased and reached a maximum soon after ($\tau\sim200$~Myr). During this time (starting at the time of the burst), material started to slowly gravitate towards the nucleus \citep[as proposed by][]{forster03}. This withdrawal affected the outermost regions first and continued inwards, causing the SFR to drop gradually, as fewer and fewer regions were forming stars intensely. This process continued until $\tau\sim50$~Myr, when disk activity dropped to a similar level as before the starburst. The nucleus may have entered an era of high SF along with the disk and maintained this high SFR throughout, to the present day, fed by the continuous supply of material, channelled in by the bar.

\subsection{Cluster Radial velocities}\label{sec:discussion:rv}
We present the measured cluster radial velocities (RVs; derived by cross-correlation with the best fitting spectroscopic models) and line-of-sight neutral interstellar gas RVs (derived from the Na\,I~D doublet line at $5890/5895$~\AA) in Fig.~\ref{plot:rot-curve}. For comparison, we have also plotted the major axis RV measurements derived from the near-IR Ca\two\ stellar absorption and Pa(10) and [S\three] nebular emission lines taken from the literature \citep{mckeith93,greve04}.

\begin{figure*}
\begin{center}
\plotone{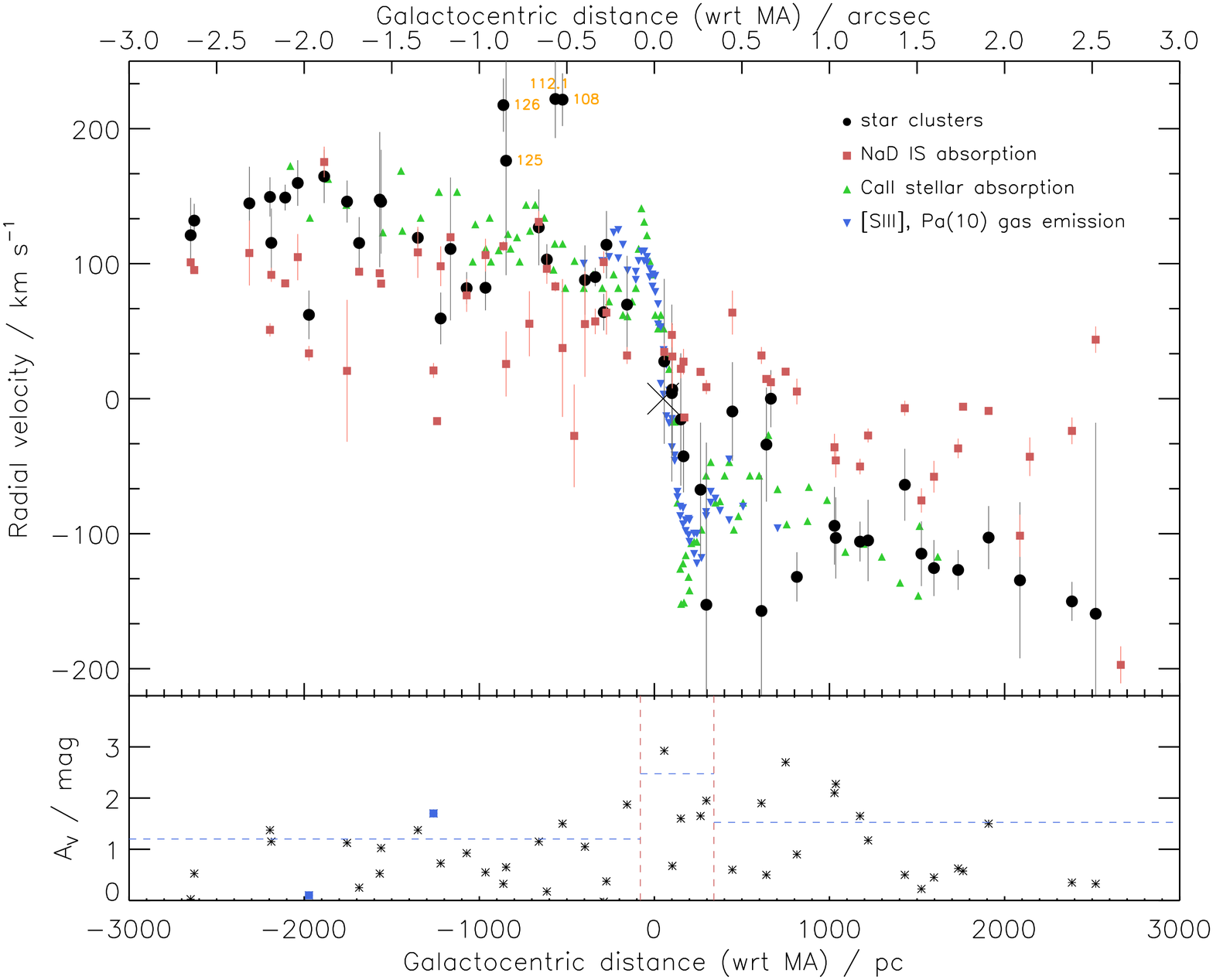}
\caption{{\bf Top panel}: Cluster radial velocities (solid circles) plotted over the major axis rotation curve defined by the near-IR Ca\two\ stellar absorption (upward-facing triangles) and [S\three] and Pa(10) nebular emission measurements (downward-facing triangles) from \citet{mckeith93} and \citet{greve04}. The cross indicates the position of the dynamical centre of M82. Rectangles represent the radial velocity of the Na\,I~D doublet  absorption line (tracing neutral interstellar gas) measured in each cluster spectrum. As these measurements do not require a high S/N, we have obtained a $v_{\rm Na\,I~D}$ for every slit in the sample, leading to Na\,I~D data with no associated cluster velocities (albeit with larger errors).
{\bf Bottom panel}: the reddening distribution of the spectroscopic cluster sample, as derived using broadband colours, and corrected for foreground emission from the MW. The dashed vertical lines roughly outline the nuclear region and the horizontal lines indicate the median extinction value for each region. The blue boxes represent the OPC candidates.
}
\label{plot:rot-curve}
\end{center}
\end{figure*}

Our observations extend the major axis coverage of RV measurements significantly further than previous studies, although we caution that our data are not strictly `major axis' measurements since our sample includes clusters located both above and below the disk mid-plane, at heights up to $\sim500$~pc. The rotation curve, as traced by the Ca\two, Pa(10) and [S\three] data, is a composite of a steep inner section tracing the nuclear bar \citep{wills00, greve04, westmoquette07c}, two flat (or gently rising) sections tracing the main disk (hereafter the flat part of the curve), and an intermediate transition region. 

In general, the cluster RVs are consistent with the Ca\two\ measurements throughout the disk. This is in itself quite interesting considering the large range in heights above and below the disk sampled by our dataset. There are, of course, a number of notable exceptions: the off-disk cluster 152 (at~$\sim-2000$~pc) has a significantly lower RV than the stellar content and is discussed below as a candidate old populous cluster. Clusters 108, 112.2, 125 and 126, located between $-500$ and $-900$~pc, have RVs 60--100~\kms\ higher than the corresponding Ca\two\ measurements, and neighbouring clusters. Besides their discrepant RVs, this group of clusters have no other distinguishing characteristics (their ages and colours are consistent with those of their neighbours), however, they are associated with region B. Their high RVs therefore support the hypothesis that region B represents a window of lower-than-average extinction, since if the clusters are located deeper within the galaxy, we would expect the radial component of their orbital velocity to be larger than that of material nearer the surface. We note that the discrepancy between the Ca\two\ and cluster RVs in region B implies that the Ca\two\ absorption originates principally from the field stars (or their scattered light), not from the star clusters.

Our observations include three slits in the nuclear region, from which we have extracted six high S/N spectra. These probably represent bright `cluster complexes' containing blended light from a number of sources\footnote{As noted before, it is not possible to deduce whether these are physical associations or chance projections}. This is because our spatial resolution is not high enough to identify individual clusters within the crowded nuclear region. The RVs of five of these six cluster complexes place them on the steep part of the rotation curve, implying that they are associated with the nuclear bar. The exact location of these clusters with respect to the galaxy nucleus was briefly discussed in \S~\ref{sec:discussion:rv}; the evidence presented by the radial velocities suggest that they may not lie at the very deepest part of the galaxy nucleus -- as also suggested by their relatively low extinction -- therefore their characterisation as `nuclear' may be inaccurate. Having said that, they provide the closest tracer to the behaviour and state of the nuclear region in our dataset. In order to study the true nuclear cluster population of this galaxy, IR observations are necessary.

Interestingly, we find that, according to the Na\,I~D velocity measurements, the neutral gas does not appear to follow the same rotation pattern as the stars, clusters or ionized gas. Throughout the disk, the Na\,I~D measurements are well described by a single, flat rotation curve with a gradient approximately equal to that of the stars/clusters in the outer disk. Na\,I~D absorption must therefore originate from neutral gas located further out in the disk (given that they absorb cluster light), where the radial component of its orbital velocity is less. This also explains why none of the Na\,I~D measurements are associated with the fast bar orbits in the nuclear region. 

\subsection{Candidate Old Populous Clusters}\label{sec:gccol}
We now return to the two discussed OPC candidates. As discussed above, cluster 152 was found to have a lower-than-expected RV for its position, with a measurement more consistent with the neutral gas RVs, implying that it may lie at a considerable distance away from the disk. No certain RV measurement could be performed for the other OPC candidate, no.~139, due to the low S/N of its spectrum (however a low-confidence fit places the cluster within the M82 system).

Both clusters are located at projected heights below the disk of $>200$~pc, implying that they may be part of the galaxy halo. A brief inspection of the area surrounding these clusters reveals the presence of numerous red stars, possibly red giants (RGs), or other,  highly reddened stars. We investigated the possibility of photometric contamination by varying the aperture/sky annulus size and found no evidence of such an effect.

Given the red colour and faintness of cluster 139 compared to the majority of clusters in the sample, there is a possibility that this object may in fact be a background galaxy -- although we have no RV data to confirm or refute this. We can reject the possibility of cluster 152 being a background galaxy (regardless of its photometry), based simply on its radial velocity, which places it within the M82 system. 
In \S~\ref{sec:ages-rvs} we found evidence that this cluster is older than $\sim8$~Gyr (using a $\frac{1}{5}\Zsun$ fit). Given the uncertainty of the Padova isochrones for intermediate/old ages, we assume that this cluster is an old Globular. However, the possibility still exists that cluster 152 is of intermediate age, constituting the first potential detection of the sort in M82. In any case, this object deems further study.

\section{Projection effects: a proposed toy model for the invisible physical structure of M82}\label{sec:model}
A careful study of all the results we have presented in this paper can reveal many pieces of information on the physical structure of the galaxy, which is obscured in our nearly edge-on viewpoint. {Such results have been presented in the past, but they have never been at the focus of any study}. It has been discussed that M82 is a spiral galaxy (as far back as OM78), with a nuclear bar spanning some 170~pc in projection \citep{achtermann95,wills00}, {which extends along our line of sight. Observations in H$\alpha$ and in the IR have also allowed for a detailed study of the supergalactic wind emanating from the galaxy nucleus, fuelled by the nuclear clusters \citep{westmoquette07c, msw09}. In this section we will present a model for the structure of the galaxy, based on the mentioned observations and the ones presented in this paper.

\subsection{Modelling disk kinematics and matching observed velocities}
Our results find a distribution of radial velocities typical of a disk galaxy, with solid body rotation in the nucleus (slits 67 and 78) and flat (or gently outwards increasing) components for disk velocities. There are, however, a handful of clusters in region~M82-B (200 to 1000~pc east of the nucleus) whose velocities do not follow the observed trend: clusters 108, 112.2, 125 and 126 all have significantly higher velocities than their neighbours and the underlying stellar component. A simple consideration of M82 as it would be seen face on reveals that clusters situated at similar projected galactocentric distance may reside at a range of depths along the line of sight. A representation of this effect may be found in Fig.~\ref{plot:rv-field}, left panel, where we have assumed that the galaxy disk is roughly circular. In this schematic, the thick line represents the nuclear bar, and crosses map out a grid limited to a 3~kpc radius (the projected visible radius of M82); vectors represent the expected radial velocity of an object in differential rotation at that location. We note here that this is a simplified representation of the galaxy, where the intricate dynamics of the regions near the bar are not modelled, and only regular, circular disk rotation is considered. This is simply because there is very little information to be revealed by nuclear clusters concerning galaxy kinematics. 

\begin{figure*}
\begin{center}
\plottwo{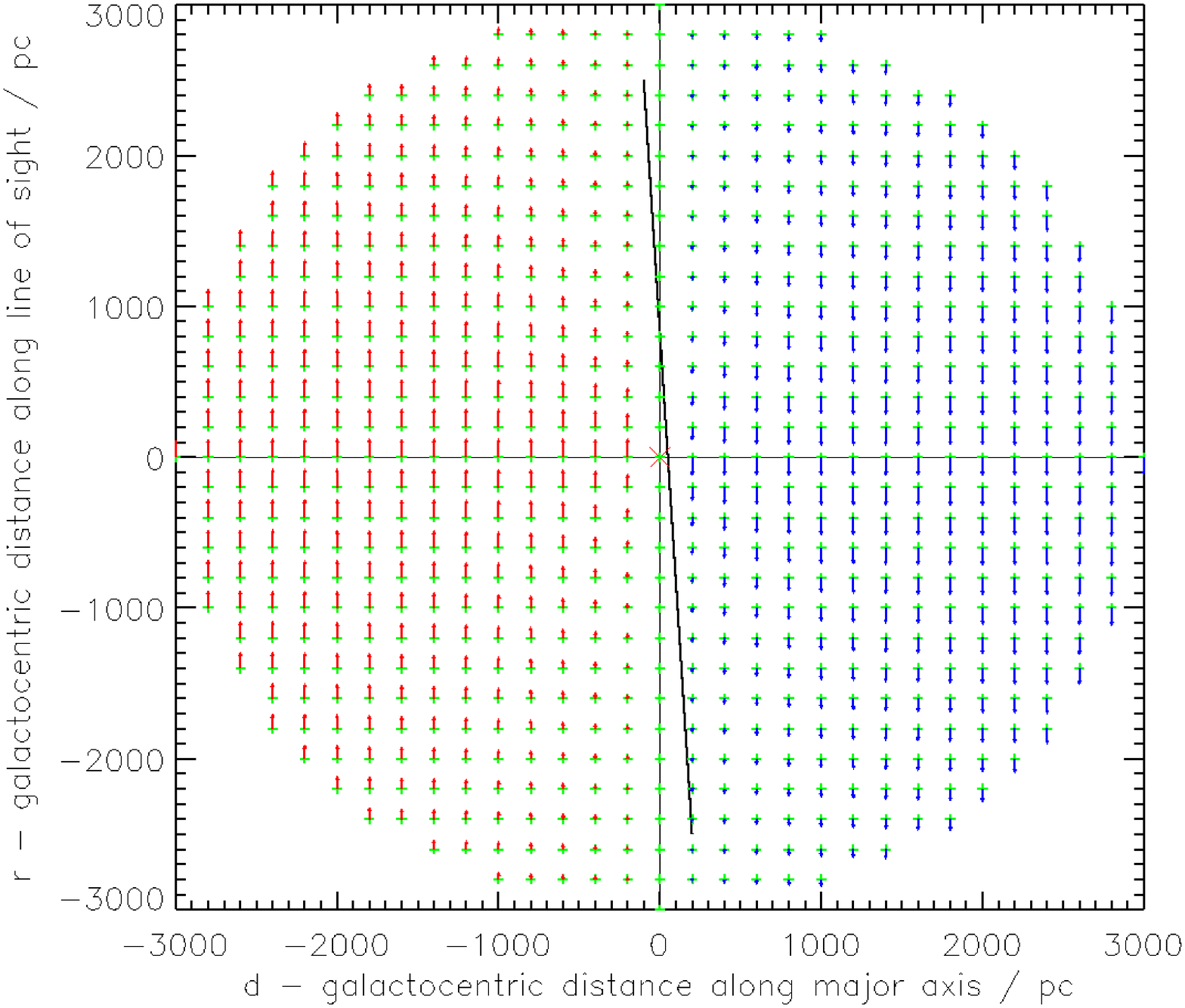}{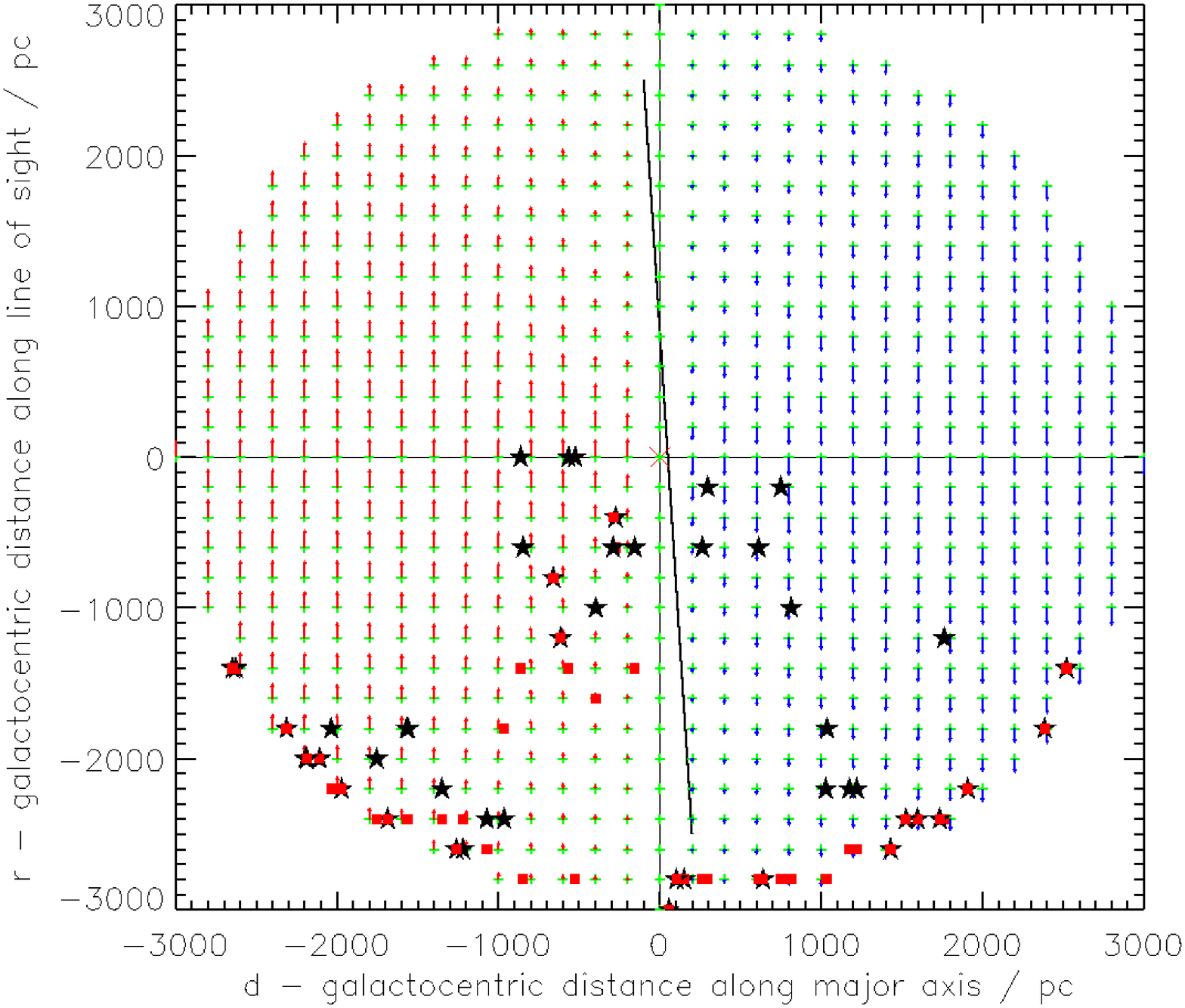}
\caption{{\bf Left}: A model of the M82 RV-field. The orientation is east to the left, and north out of the page (following our observed viewpoint of M82). Crosses represent grid positions with galactocentric distance less than 3~kpc, which we assume to be the full extent of the (hypothetically circular) M82 disk. The thick line down the centre of the plot represents the approximate location of the bar (the extremes on the $r$-axis are uncertain, as we can only observe its projected location). Vectors represent modelled radial velocities, and are colour-coded as receding (red, east side) and approaching (blue, west side). The grid locations with plotted velocities are all outside the inner, solid-body rotation region associated with the bar (represented by the thick tilted dark line). The observable radial component of this velocity ranges from zero at $d=0$, where the velocity is tangential to our line of sight, to the maximum velocity on the $r=0$ line (where all of the motion is in the radial direction). 
{\bf Right:} The same model, with overplotted predicted cluster locations (stars) based on the kinematics of the cluster and nearest pocket of interstellar gas (Na\,I~D, red squares). This is not an accurate depiction, rather a diagnostic to confirm our hypothesis on the location of clusters in region~B, where this demonstrates the depth at which these clusters are located. 
}
\label{plot:rv-field}
\end{center}
\end{figure*}

This basic trigonometric consideration reveals that a cluster on a regular orbit near the visible surface of the galaxy will appear to move significantly slower than another regularly orbiting cluster situated at great depth, $r$. The weight of this effect ranges with projected galactocentric distance (indicated by $d$ in Fig.~\ref{plot:rv-field}): at the extremes of the galaxy disk this difference will be within the measurement  errors;  in the mid-regions of the galaxy, assuming the M82 disk is roughly circular, the velocities will differ by a factor of two. Clusters projected on top of the nucleus will appear to have near-zero RVs, as their velocity is mostly in the tangential direction and is therefore not measureable; just to the side of the nucleus the discrepancy reaches a factor of 10.

\subsection{Spiral structure}
Any existing spiral structure in M82 will affect its appearance, even in our highly inclined viewpoint. Given the orientation of the bar, which extends along our line of sight, one would expect spiral arms to originate from the visible central regions of the galaxy on the near side, and a corresponding site in the obscured side. This could result in a highly irregular dust distribution, and therefore significant discrepancies in the visible components of the disk.

In order to find evidence of the two-armed, barred spiral structure \citep[as hypothesised in the literature, see][and references therein]{yun94} one must also consider the inclination of the disk, which provides a view of M82 from below. Assuming no significant spiral sub-structure other than the two dominant arms, we should expect to face one strong extinction front in the one side of the galaxy (the near spiral arm), and an `opening' in the dust distribution on the other side, along lines of sight underneath the bulk of the galaxy, and into the disk. More specifically, the near spiral arm should stem from the tip of the bar \citep[the approximate location of SSC A1][]{ljs06} and extend westwards, and the far-side arm should originate at a symmetric location and `wrap' around region~B. Such a conclusion was reached by \citet{mayya05}, who derived the approximate location of the spiral arms by subtracting a hypothetical axisymmetric exponential disk from NIR imaging of the galaxy.

This structure is observed in the optical and our hypothesis on the approximate placing of the two spiral arms is supported by long wavelength observations (IR and radio). Examining the optical data presented in this paper, we assign the proposed structure as follows: the western side of the galaxy hosts the near spiral arm, hence the highly obscuring extinction front in the southwest disk. A faint trail of reddening dust can be seen superimposed on the central region, stemming from the general vicinity of tip of the nuclear bar, and spreading towards this extinction front. Therefore, the eastern side is expected to host the far spiral arm; as it hosts region~B, where kinematics reveal lines of sight deep into the body of the galaxy, it is consistent with this concept.

Longer wavelength studies lend additional support to this model. First, we consider Hubble {\it NICMOS}  data \citep{alonso-herrero-m82} in the $J$ and $H$ near-infrared bands, which trace star light, with much less obscuration from interstellar dust\footnote{Central wavelengths are $1.10$ and $1.60\mu$m for the $J$ and $H$ bands respectively; assuming a galactic extinction law, the exact relation is $A_J=0.3$ and $A_H=0.1$~mag for each magnitude of extinction in $V$.}. These data show the eastern disk as a crowded stellar field, with a large number of star clusters that are invisible in sub-micron wavebands. The part of region~B closest to the nucleus (200 to 700~pc) is more crowded, as it represents a cross section of more than $\sim4$~kpc in depth, and projections of stars and clusters are the norm. Unfortunately, no equivalent high-resolution IR data exist for the western disk. 

As a tracer of spiral arms we have sought out radio data with continuous coverage of the galaxy. We use the CO molecular data in \citet[][see their Fig.~2 where radio and optical data are overlayed]{walter02} as an appropriate diagnostic of spiral structure -- CO is a tracer of H$_2$. We find a good agreement between the structures we propose as representing the spiral arms and the CO streamers discovered by \citeauthor{walter02}. The radial velocities found for these streamers are consistent with material delineating the outer parts of the disk, as a spiral arm would. 

This model finds support in our analysis of disk kinematics; in the right panel of Fig.~\ref{plot:rv-field} we have matched observed cluster velocities onto the modelled RV field, according to their measured projected galactocentric distance ($d$). Applying this method, we find clusters in region~B to be consistent with being located at large radii ($r$), given their high radial velocities. 

Finally, we consider the placement of LOS gas, traced by the Na\,I~D doublet line (plotted onto Fig.~\ref{plot:rv-field} as filled red boxes). We expect Na\,I~D line velocities to have maxima at extreme $d$ (i.~e. the western and eastern edges of the disk), and a linear trend between those extrema crossing $(d,r)=(0,0)$; this is confirmed by our observations. There is one deviation from this, as we find the interstellar gas to be situated at greater depth across region~B (but at the surface between $r=600-800$), thus lending some more support to our hypothesis. 

Na\,I~D measurements can also provide more insight on the exact location of their corresponding clusters. As it is not possible to distinguish kinematically between symmetric near/far side locations, we resort to indirect means to choose between the two solutions to this problem: in some cases, the Na\,I~D clump is placed behind the respective cluster ({\it wrt} our LOS), which is impossible as the line occurs in absorption of cluster light. In these cases we deduce a far side location for the cluster.

We therefore present new observational evidence for the morphological structure of M82, and confirm that it consists of two dominant spiral arms and very weak, if any, other spiral structure, as deduced by \citet{ichikawa95} using NIR imaging of the galaxy, and followed up by \citet{mayya05}.

\subsection{Implications of the proposed model on region M82-B}
Under the proposed model, part of the eastern M82 disk is left relatively free of extinction. The spiral arm is situated at the far reaches of the galaxy, therefore allowing a clear view into the body of the galaxy. In Paper~I and \citet{smith07regb} we argued that region B in the eastern disk is seen through low extinction windows, similar to ``Baade's windows'' in the Milky Way. Here, we offer the interpretation that these windows form as a result of the inclination of M82 and provide  a line of sight under the spiral arm extinction front. Thus, we confirm that region B is not intrinsically brighter, but simply seen under a clearer, relatively unextinguished line of sight. This interpretation is supported by the high radial velocities of the mentioned group of region~B clusters, as they lie within these low extinction windows. This means that they can be seen clearly, even though they are situated  deep within the body of the galaxy, and they do not form a distinct population, in terms of SFR, intrinsic luminosity, or actual kinematics.

\section{Summary}
In this paper, we have presented a spectroscopic and photometric study of the disk cluster population of M82, using a sample of 44 bright, isolated disk clusters. We have also studied five cluster complexes in the galaxy nucleus, which, together with the 44 disk clusters, provide virtually continuous coverage across the galaxy and comprise the largest to date spectroscopic sample of extragalactic young clusters. 

Our primary goal was to derive spectroscopic (age, radial velocity) and photometric (colour, extinction, mass) information on the clusters, and to characterise the population based on the properties of the individuals. In addition, we have extracted some information on the environment which they inhabit, in terms of the M82 starburst history. 

Having combined information from spectroscopy and imaging, we found the disk clusters  to form a uniform population, displaying no positional dependencies with respect to age. We have therefore roughly derived the starburst history of M82, based on the cluster age distribution: this we describe as a long period of low SF activity, followed by a burst of star formation that gradually decreased in the disk, while the nucleus has been starbursting for at least the last 15~Myr, and perhaps throughout the starburst, and appears to be leading SF in the galaxy at the current epoch.

{\it In summary}, we conclude the following:

\begin{itemize}

\item The disk cluster population of M82 is consistent with having formed during the last encounter with M81, \mbox{$\sim220$~Myr} ago. This is based on spectroscopic ages, which range between $(30-270)$~Myr.

\item The SFR of M82 reached a peak some $\sim150$~Myr ago, and has remained virtually constant since, with the disk starburst leading star/cluster formation up to $\tau\sim50$~Myr, and the nuclear starburst taking over since.

\item The nuclear starburst appears to be ongoing, as indicated by the possible detection of WR features in the central regions of the galaxy.

\item The only possible evidence of pre-starburst cluster formation comes from cluster 152, a candidate  old populous / Globular cluster, that was observed for the first time.

\item Cluster reddening ranges between $0.4-3.2$~mag in the disk and $1.4-3.5$~mag in the nuclear region. The measured masses range between typical low-end values of $\sim10^4$~\Msun and reach $\sim10^6$~\Msun. From these values, a total mass of $\sim10^8$ is deduced for the star cluster population of M82.

\item The cluster system follows a regular rotation pattern about the nucleus, with a nearly flat rotation curve component for the disk, and a solid-body rotation part in the bar area. 

\item Region~M82-B is a regular part of the galaxy disk, and represents a line of sight into the body of the galaxy through lower average extinction windows, made possible by the inclination and spiral structure of the galaxy.

\item The data presented in this work support previous predictions of a two spiral arm system, as indicated by our  study of cluster radial velocities. We place the roots of the near spiral arm in the vicinity of SSC A1 (the tip of the bar) extending west, and the end of the far-side arm above region~B in our subtended viewpoint.

\end{itemize}

This concludes the spectroscopic and photometric study of the M82 cluster population we initiated in \citet{smith07regb}. This series has included, apart from Paper I, investigations of the environment of the M82 starburst core \citep{westmoquette07c} and the dynamics of the disk and inner wind \citep{msw09}. 

\acknowledgments{
We would like to thank the anonymous referee for insightful suggestions that led to an improved manuscript. ISK would like to thank Mikako Matsuura for useful discussions on the CO content of M82. JSG's research was partially funded by the National Science Foundation through grant AST-0708967 to the University of Wisconsin-Madison.
}

\bibliographystyle{apj}
\defcitealias{isk08a}{Paper I}
\bibliography{references}

\begin{thebibliography}{52}
\expandafter\ifx\csname natexlab\endcsname\relax\def\natexlab#1{#1}\fi

\bibitem[{{Achtermann} \& {Lacy}(1995)}]{achtermann95}
{Achtermann}, J.~M., \& {Lacy}, J.~H. 1995, ApJ, 439, 163

\bibitem[{{Alonso-Herrero} {et~al.}(2003){Alonso-Herrero}, {Rieke}, {Rieke}, \&
  {Kelly}}]{alonso03}
{Alonso-Herrero}, A., {Rieke}, G.~H., {Rieke}, M.~J., \& {Kelly}, D.~M. 2003,
  AJ, 125, 1210

\bibitem[{{Alonso-Herrero} {et~al.}(2001){Alonso-Herrero}, {Rieke}, {Rieke}, \&
  {Kelly}}]{alonso-herrero-m82}
{Alonso-Herrero}, A., {Rieke}, M.~J., {Rieke}, G.~H., \& {Kelly}, D.~M. 2001,
  AP\&SS, 276, 1109

\bibitem[{{Anders} {et~al.}(2004){Anders}, {Bissantz}, {Fritze-v.~Alvensleben},
  \& {de Grijs}}]{anders04}
{Anders}, P., {Bissantz}, N., {Fritze-v.~Alvensleben}, U., \& {de Grijs}, R.
  2004, MNRAS, 347, 196

\bibitem[{{Anders} \& {Fritze-v.~Alvensleben}(2003)}]{anders03-galev}
{Anders}, P., \& {Fritze-v.~Alvensleben}, U. 2003, A\&A, 401, 1063

\bibitem[{{Barker} {et~al.}(2008){Barker}, {de Grijs}, \&
  {Cervi{\~n}o}}]{barker08}
{Barker}, S., {de Grijs}, R., \& {Cervi{\~n}o}, M. 2008, A\&A, 484, 711

\bibitem[{{Bastian}(2008)}]{bastian08sfr}
{Bastian}, N. 2008, MNRAS, 390, 759

\bibitem[{{Bastian} {et~al.}(2006){Bastian}, {Emsellem}, {Kissler-Patig}, \&
  {Maraston}}]{bastian06complex}
{Bastian}, N., {Emsellem}, E., {Kissler-Patig}, M., \& {Maraston}, C. 2006,
  A\&A, 445, 471

\bibitem[{{Bastian} {et~al.}(2008){Bastian}, {Gieles}, {Goodwin}, {Trancho},
  {Smith}, {Konstantopoulos}, \& {Efremov}}]{bastian08cores}
{Bastian}, N., {Gieles}, M., {Goodwin}, S.~P., {Trancho}, G., {Smith}, L.~J.,
  {Konstantopoulos}, I., \& {Efremov}, Y. 2008, MNRAS, 389, 223

\bibitem[{{Bastian} {et~al.}(2005){Bastian}, {Gieles}, {Lamers}, {Scheepmaker},
  \& {de Grijs}}]{bastian05a}
{Bastian}, N., {Gieles}, M., {Lamers}, H.~J.~G.~L.~M., {Scheepmaker}, R.~A., \&
  {de Grijs}, R. 2005, A\&A, 431, 905

\bibitem[{{Bastian} {et~al.}(2007){Bastian}, {Konstantopoulos}, {Smith},
  {Trancho}, {Westmoquette}, \& {Gallagher}}]{bastian07m82f}
{Bastian}, N., {Konstantopoulos}, I., {Smith}, L.~J., {Trancho}, G.,
  {Westmoquette}, M.~S., \& {Gallagher}, J.~S. 2007, MNRAS, 379, 1333

\bibitem[{{Bibby} {et~al.}(2008){Bibby}, {Crowther}, {Furness}, \&
  {Clark}}]{bibby08}
{Bibby}, J.~L., {Crowther}, P.~A., {Furness}, J.~P., \& {Clark}, J.~S. 2008,
  MNRAS, 386, L23

\bibitem[{{Bruzual} \& {Charlot}(2003)}]{bc03}
{Bruzual}, G., \& {Charlot}, S. 2003, MNRAS, 344, 1000

\bibitem[{{de Grijs} {et~al.}(2003){de Grijs}, {Bastian}, \& {Lamers}}]{RdG03a}
{de Grijs}, R., {Bastian}, N., \& {Lamers}, H.~J.~G.~L.~M. 2003, MNRAS, 340,
  197

\bibitem[{{de Grijs} {et~al.}(2001){de Grijs}, {O'Connell}, \&
  {Gallagher}}]{RdG01}
{de Grijs}, R., {O'Connell}, R.~W., \& {Gallagher}, III, J.~S. 2001, AJ, 121,
  768

\bibitem[{{Filippenko}(1982)}]{filippenko82}
{Filippenko}, A.~V. 1982, PASP, 94, 715

\bibitem[{{F{\"o}rster Schreiber} {et~al.}(2003){F{\"o}rster Schreiber},
  {Genzel}, {Lutz}, \& {Sternberg}}]{forster03}
{F{\"o}rster Schreiber}, N.~M., {Genzel}, R., {Lutz}, D., \& {Sternberg}, A.
  2003, ApJ, 599, 193

\bibitem[{{Gallagher} \& {Smith}(1999)}]{GnS}
{Gallagher}, J.~S., \& {Smith}, L.~J. 1999, MNRAS, 304, 540

\bibitem[{{Gieles}(2009)}]{gieles-cimf}
{Gieles}, M. 2009, MNRAS, 277

\bibitem[{{Gieles} {et~al.}(2005){Gieles}, {Bastian}, {Lamers}, \&
  {Mout}}]{gieles05}
{Gieles}, M., {Bastian}, N., {Lamers}, H.~J.~G.~L.~M., \& {Mout}, J.~N. 2005,
  A\&A, 441, 949

\bibitem[{{Gieles} {et~al.}(2007){Gieles}, {Lamers}, \& {Portegies
  Zwart}}]{gieles07b}
{Gieles}, M., {Lamers}, H.~J.~G.~L.~M., \& {Portegies Zwart}, S.~F. 2007, ApJ,
  668, 268

\bibitem[{{Gonz{\'a}lez-Delgado} {et~al.}(2005){Gonz{\'a}lez-Delgado},
  {Cervi{\~n}o}, {Martins}, {Leitherer}, \& {Hauschildt}}]{gd05}
{Gonz{\'a}lez-Delgado}, R.~M., {Cervi{\~n}o}, M., {Martins}, L.~P.,
  {Leitherer}, C., \& {Hauschildt}, P.~H. 2005, MNRAS, 357, 945

\bibitem[{{Greve}(2004)}]{greve04}
{Greve}, A. 2004, A\&A, 416, 67

\bibitem[{{Ichikawa} {et~al.}(1995){Ichikawa}, {Yanagisawa}, {Itoh},
  {Tarusawa}, {van Driel}, \& {Ueno}}]{ichikawa95}
{Ichikawa}, T., {Yanagisawa}, K., {Itoh}, N., {Tarusawa}, K., {van Driel}, W.,
  \& {Ueno}, M. 1995, AJ, 109, 2038

\bibitem[{{Konstantopoulos} {et~al.}(2008){Konstantopoulos}, {Bastian},
  {Smith}, {Trancho}, {Westmoquette}, \& {Gallagher}}]{isk08a}
{Konstantopoulos}, I.~S., {Bastian}, N., {Smith}, L.~J., {Trancho}, G.,
  {Westmoquette}, M.~S., \& {Gallagher}, III, J.~S. 2008, ApJ, 674, 846, {Paper
  I}

\bibitem[{{Lamers} {et~al.}(2005){Lamers}, {Gieles}, {Bastian}, {Baumgardt},
  {Kharchenko}, \& {Portegies Zwart}}]{lamers05}
{Lamers}, H.~J.~G.~L.~M., {Gieles}, M., {Bastian}, N., {Baumgardt}, H.,
  {Kharchenko}, N.~V., \& {Portegies Zwart}, S. 2005, A\&A, 441, 117

\bibitem[{{Lampton} {et~al.}(1976){Lampton}, {Margon}, \& {Bowyer}}]{lampton76}
{Lampton}, M., {Margon}, B., \& {Bowyer}, S. 1976, ApJ, 208, 177

\bibitem[{{Larsen}(2009)}]{larsen09}
{Larsen}, S.~S. 2009, A\&A, 494, 539

\bibitem[{{Mayya} {et~al.}(2006){Mayya}, {Bressan}, {Carrasco}, \&
  {Hernandez-Martinez}}]{mayya06}
{Mayya}, Y.~D., {Bressan}, A., {Carrasco}, L., \& {Hernandez-Martinez}, L.
  2006, ApJ, 649, 172

\bibitem[{{Mayya} {et~al.}(2005){Mayya}, {Carrasco}, \& {Luna}}]{mayya05}
{Mayya}, Y.~D., {Carrasco}, L., \& {Luna}, A. 2005, ApJl, 628, L33

\bibitem[{{McKeith} {et~al.}(1993){McKeith}, {Castles}, {Greve}, \&
  {Downes}}]{mckeith93}
{McKeith}, C.~D., {Castles}, J., {Greve}, A., \& {Downes}, D. 1993, A\&A, 272,
  98

\bibitem[{{McLeod} {et~al.}(1993){McLeod}, {Rieke}, {Rieke}, \&
  {Kelly}}]{mcleod93}
{McLeod}, K.~K., {Rieke}, G.~H., {Rieke}, M.~J., \& {Kelly}, D.~M. 1993, ApJ,
  412, 111

\bibitem[{{McQuinn} {et~al.}(2009){McQuinn}, {Skillman}, {Cannon}, {Dalcanton},
  {Dolphin}, {Stark}, \& {Weisz}}]{mcquinn09}
{McQuinn}, K.~B.~W., {Skillman}, E.~D., {Cannon}, J.~M., {Dalcanton}, J.~J.,
  {Dolphin}, A., {Stark}, D., \& {Weisz}, D. 2009, ApJ, 695, 561

\bibitem[{{Melo} {et~al.}(2005){Melo}, {Mu{\~n}oz-Tu{\~n}{\'o}n},
  {Ma{\'{\i}}z-Apell{\'a}niz}, \& {Tenorio-Tagle}}]{melo05}
{Melo}, V.~P., {Mu{\~n}oz-Tu{\~n}{\'o}n}, C., {Ma{\'{\i}}z-Apell{\'a}niz}, J.,
  \& {Tenorio-Tagle}, G. 2005, ApJ, 619, 270

\bibitem[{{Mutchler} {et~al.}(2007){Mutchler}, {Bond}, {Christian}, {Frattare},
  {Hamilton}, {Januszewski}, {Levay}, {Mountain}, {Noll}, {Royle}, {Gallagher},
  \& {Puxley}}]{mutchler07}
{Mutchler}, M., {et~al.} 2007, PASP, 119, 1

\bibitem[{{O'Connell} {et~al.}(1995){O'Connell}, {Gallagher}, {Hunter}, \&
  {Colley}}]{oconnell95}
{O'Connell}, R.~W., {Gallagher}, III, J.~S., {Hunter}, D.~A., \& {Colley},
  W.~N. 1995, ApJL, 446, L1+

\bibitem[{{O'Connell} \& {Mangano}(1978)}]{om78}
{O'Connell}, R.~W., \& {Mangano}, J.~J. 1978, ApJ, 221, 62

\bibitem[{{Rieke} \& {Lebofsky}(1985)}]{rieke}
{Rieke}, G.~H., \& {Lebofsky}, M.~J. 1985, ApJ, 288, 618

\bibitem[{{Saito} {et~al.}(2005){Saito}, {Ohyama}, {Yoshida}, {Sasaki},
  {Kosugi}, {Kashikawa}, {Takata}, {Shimizu}, {Inata}, {Okita}, {Aoki},
  {Sekiguchi}, {Kawabata}, {Asai}, {Taguchi}, {Ebizuka}, {Yadoumaru}, {Ozawa},
  \& {Iye}}]{saito05}
{Saito}, Y., {et~al.} 2005, ApJ, 621, 750

\bibitem[{{Satyapal} {et~al.}(1995){Satyapal}, {Watson}, {Pipher}, {Forrest},
  {Coppenbarger}, {Raines}, {Libonate}, {Piche}, {Greenhouse}, {Smith},
  {Thompson}, {Fischer}, {Woodward}, \& {Hodge}}]{satyapal95}
{Satyapal}, S., {et~al.} 1995, ApJ, 448, 611

\bibitem[{{Schlegel} {et~al.}(1998){Schlegel}, {Finkbeiner}, \&
  {Davis}}]{schlegel98}
{Schlegel}, D.~J., {Finkbeiner}, D.~P., \& {Davis}, M. 1998, ApJ, 500, 525

\bibitem[{{Sidoli} {et~al.}(2006){Sidoli}, {Smith}, \& {Crowther}}]{sidoli06}
{Sidoli}, F., {Smith}, L.~J., \& {Crowther}, P.~A. 2006, MNRAS, 370, 799

\bibitem[{{Smith} \& {Gallagher}(2001)}]{SnG}
{Smith}, L.~J., \& {Gallagher}, J.~S. 2001, MNRAS, 326, 1027

\bibitem[{{Smith} {et~al.}(2006){Smith}, {Westmoquette}, {Gallagher},
  {O'Connell}, {Rosario}, \& {de Grijs}}]{ljs06}
{Smith}, L.~J., {Westmoquette}, M.~S., {Gallagher}, J.~S., {O'Connell}, R.~W.,
  {Rosario}, D.~J., \& {de Grijs}, R. 2006, MNRAS, 370, 513

\bibitem[{{Smith} {et~al.}(2007){Smith}, {Bastian}, {Konstantopoulos},
  {Gallagher}, {Gieles}, {de Grijs}, {Larsen}, {O'Connell}, \&
  {Westmoquette}}]{smith07regb}
{Smith}, L.~J., {et~al.} 2007, ApJL, 667, L145

\bibitem[{{Trancho} {et~al.}(2007){Trancho}, {Bastian}, {Schweizer}, \&
  {Miller}}]{gelys07a}
{Trancho}, G., {Bastian}, N., {Schweizer}, F., \& {Miller}, B.~W. 2007, ApJ,
  658, 993

\bibitem[{{Walter} {et~al.}(2002){Walter}, {Weiss}, \& {Scoville}}]{walter02}
{Walter}, F., {Weiss}, A., \& {Scoville}, N. 2002, ApJl, 580, L21

\bibitem[{{Westmoquette} {et~al.}(2009){Westmoquette}, {Smith}, {Gallagher},
  {Trancho}, {Bastian}, \& {Konstantopoulos}}]{msw09}
{Westmoquette}, M.~S., {Smith}, L.~J., {Gallagher}, J.~S., {Trancho}, G.,
  {Bastian}, N., \& {Konstantopoulos}, I.~S. 2009, ApJ, 696, 192

\bibitem[{{Westmoquette} {et~al.}(2007){Westmoquette}, {Smith}, {Gallagher},
  {O'Connell}, {Rosario}, \& {de Grijs}}]{westmoquette07c}
{Westmoquette}, M.~S., {Smith}, L.~J., {Gallagher}, III, J.~S., {O'Connell},
  R.~W., {Rosario}, D.~J., \& {de Grijs}, R. 2007, ApJ, 671, 358

\bibitem[{{Wills} {et~al.}(2000){Wills}, {Das}, {Pedlar}, {Muxlow}, \&
  {Robinson}}]{wills00}
{Wills}, K.~A., {Das}, M., {Pedlar}, A., {Muxlow}, T.~W.~B., \& {Robinson},
  T.~G. 2000, MNRAS, 316, 33

\bibitem[{{Yun}(1999)}]{yun99}
{Yun}, M.~S. 1999, in IAU Symposium, Vol. 186, Galaxy Interactions at Low and
  High Redshift, ed. J.~E. {Barnes} \& D.~B. {Sanders}, 81

\bibitem[{{Yun} {et~al.}(1994){Yun}, {Ho}, \& {Lo}}]{yun94}
{Yun}, M.~S., {Ho}, P.~T.~P., \& {Lo}, K.~Y. 1994, Nature, 372, 530

\end{thebibliography}

\end{document}